\documentclass[aps,showpacs,preprintnumbers,amsmath,notitlepage,nofootinbib,amssymb,longbibliography]{revtex4-2}
\usepackage[english]{babel}
\usepackage{graphicx}  
\usepackage{dcolumn}   
\usepackage{bm}         
\usepackage{float}
\usepackage{multirow}
\usepackage{diagbox}
\usepackage{xcolor}
\usepackage{tikz}
\usepackage{tikzscale}
\usepackage{neuralnetwork}
\usepackage{comment} 
\usetikzlibrary{arrows,backgrounds}
\usepgflibrary{shapes.multipart}
\usepackage{graphicx}  
\usepackage{dcolumn}   
\usepackage{bm}        
\usepackage{amssymb}   
\usepackage{amsmath}
\usepackage{hyperref}
\usepackage[caption=false]{subfig}
\usepackage{amsmath}
\usepackage{indentfirst}
\usepackage{array}
\usepackage{xcolor}
\usepackage{marginnote}

\usepackage[normalem]{ulem}

\begin{document}
\title{Machine learning spectral functions in lattice QCD}

\preprint{\today}
\author{S.-Y. Chen$^{\rm a}$}
\email[]{shiyang-chen@mails.ccnu.edu.cn}   
\author{H.-T. Ding$^{\rm a}$}
\author{F.-Y. Liu$^{\rm a,b}$}
\email[]{fyliu@mails.ccnu.edu.cn}
\author{G. Papp$^{\rm b}$}
\author{C.-B. Yang$^{\rm a}$}

\affiliation{
	$^{\rm a}$Key Laboratory of Quark and Lepton Physics (MOE) \& Institute of Particle Physics, \\
	Central China Normal University, Wuhan 430079, China\\
	$^{\rm b}$Institute for Physics, E{\"o}tv{\"o}s Lor\'and University, \\1/A P\'azm\'any P. s\'et\'any, H-1117, Budapest, Hungary
}

\begin{abstract}

We study the inverse problem of reconstructing spectral functions from Euclidean correlation functions via machine learning. We propose a novel neural network, SVAE, which is based on the variational autoencoder (VAE) and can be naturally applied to the inverse problem. The prominent feature of the SVAE is that a Shannon-Jaynes entropy term having the ground truth values of spectral functions as prior information is included in the loss function to be minimized. We train the network with general spectral functions produced from a Gaussian mixture model. As a test, we use correlators generated from four different types of physically motivated spectral functions made of one resonance peak, a continuum term and perturbative spectral function obtained using non-relativistic QCD. From the mock data test we find that the SVAE in most cases is comparable to the maximum entropy method (MEM) in the quality of reconstructing spectral functions and even outperforms the MEM in the case where the spectral function has sharp peaks with insufficient number of data points in the correlator. By applying to temporal correlation functions of charmonium in the pseudoscalar channel obtained in the quenched lattice QCD at 0.75 $T_c$ on $128^3\times96$ lattices and $1.5$ $T_c$ on $128^3\times48$ lattices, we find that the resonance peak of $\eta_c$ extracted from both the SVAE and MEM has a substantial dependence on the number of points in the temporal direction ($N_\tau$) adopted in the lattice simulation and $N_\tau$ larger than 48 is needed to resolve the fate of $\eta_c$ at 1.5 $T_c$.
\end{abstract}

\maketitle


\section{Introduction} 
\label{sec:intro}
Mapping out the QCD phase diagram and understanding the properties of the quark-gluon plasma (QGP) have been the central goals in the high energy nuclear physics. The light and heavy hadrons could reflect the chiral and deconfinement aspects of the rapid crossover transition from the hadron phase to the quark-gluon plasma phase. The degeneracy of light meson, e.g. $\rho$ and $a_1$, reflects the chiral symmetry restoration~\cite{Rapp:1999ej,Bazavov:2019www}, while the dissociation of quarkonia, e.g. $J/\psi$ and $\Upsilon$, which can serve as the thermometer of the QGP~\cite{Matsui:1986dk,Karsch:2005nk}, could probe the deconfinement aspect of the crossover transition. An enhancement of dilepton yields in the low invariant mass region~\cite{STAR:2013pwb,PHENIX:2015vek,ALICE:2018ael}, and a suppression of quarkonia yields~\cite{PHENIX:2014tbe,STAR:2013kwk,CMS:2017ycw,ALICE:2018wzm} in nucleus-nucleus collisions compared to those in the nucleon-nucleon collisions have been observed in RHIC and LHC energies.
One of the important ingredients to interpret the experimental observation is to understand the properties of $\rho$ and quarkonia states in the quark-gluon plasma phase~\cite{Rapp:1999ej,Aarts:2016hap,Ding:2020rtq,Rothkopf:2019ipj}. Theoretically these properties are all related to the hadron spectral functions. Thermal dilepton rates in the quark gluon plasma can be obtained directly via the spectral function of $\rho$~\cite{McLerran:1984ay}, and the fate of quarkonia in the hot medium can be determined from the disappearance of peak structures in the quarkonia spectral functions~\cite{Asakawa:2003re}. On the other hand, transport properties of QGP are also related to spectral functions. For instance the heavy quark diffusion coefficient and electrical conductivity of QGP can be determined from the slope of corresponding vector spectral function at zero frequency~\cite{Ding:2015ona}, while the viscosities of QGP can also be determined from the corresponding spectral function of energy momentum tensor~\cite{Meyer:2011gj}. Due to their non-perturbative features
it thus is important to understand e.g. the dissociation temperature of quarkonia, transport and electromagnetic properties of the hot medium from lattice QCD (LQCD) simulations. 
However, despite the importance of the spectral functions they cannot be calculated directly by using LQCD~\cite{Ding:2015ona}.

The spectral function is encoded in the Euclidean two-point correlation function which is directly accessible in LQCD. The relation between the hadron spectral function $\rho(\omega,T)$~\footnote{This shall not to be confused with the vector meson $\rho$ mentioned in the above paragraph and hereafeter $\rho$ stands for the hadron spectral function.} and the Euclidean temporal correlator $G(\tau,T)$ is expressed as follows
\begin{equation}
G(\tau,T)=\sum_{x,y,z} 
\left\langle J_H(0,\vec{0})J_H^+(\tau,\vec{x}) \right\rangle_T=\int_0^\infty \!\frac{\mathrm{d}\omega}{2\pi}\ K(\omega,\tau,T) \rho(\omega,T) \,,
\label{eq:Grho}
\end{equation}
where $\langle\cdots\rangle$ stands for the thermal average at a temperature $T$, $J_H=\bar{\psi}\Gamma_H\psi$ is a bilinear form of fermion fields $\psi$ with $H$ the label for states in different channels, e.g. $\Gamma_H=\gamma_5$ denotes the pseudoscalar channel. The summation of $\langle J_HJ_H^+\rangle$ over spatial coordinates $x$, $y$ and $z$ can be computed via LQCD simulations. The coordinate in the temporal direction, $\tau$, ranges from 0 to $1/T$. The integral kernel $K(\omega,\tau,T)$ is expressed as follows
\begin{equation}
K(\omega,\tau,T) = \frac{\cosh(\omega(\tau-\frac1{2T}))}{\sinh(\frac{\omega}{2T})}\,,
\label{eq:kernel}
\end{equation}
which is symmetric at $\tau=1/2T$ and has a trivial temperature dependence. Although this relation is straightforward, the number of points $G(\tau,T)$ in the temporal direction at a fixed temperature, $N_\tau$, is limited as $T=1/(aN_\tau)$ with $a$ the lattice spacing. Since $N_\tau$ is normally of the order of $\mathcal{O}(10)$ due to large computing costs of LQCD computations, the reconstruction of $\rho(\omega,T)$, which typically has $\mathcal{O}(1000)$ points in the frequency ($\omega$) space, is thus an ill-posed inverse problem. To circumvent such an intrinsic problem, the maximum entropy method (MEM) based on the Bayesian theorem has been introduced in the reconstruction of $\rho(\omega,T)$ from the Euclidean lattice correlators many years ago~\cite{Asakawa:2003re}. The MEM can give an unique solution of the most probable image of $\rho(\omega,T)$ if the maximum of a linear combination of the Shannon-Jaynes entropy and likelihood function exists. However, the reconstructed spectral function from the MEM depends on the prior information of the spectral function, which is the so-called default model being the crucial input of the MEM. A variant version of the MEM has also been proposed recently~\cite{Burnier:2013nla} and it implicitly keeps the default model fixed in the analysis to avoid the default model dependence~\cite{Burnier:2014ssa,Asakawa:2020hjs,Rothkopf:2020qqt}. To reduce the dependence on the prior information, stochastic methods based on the Bayesian theorem, e.g. stochastic optimization method (SOM) and stochastic analytical inference (SAI) have been introduced recently and are found to yield the consistent results with those from the standard MEM~\cite{Ding:2017std}. Some other methods, e.g., the Backus Gilbert method~\cite{Backus:1968} and Tikhonov regularization~\cite{Tikhonov:1963} are also used in the full QCD simulations where the number of points in the temporal direction is relatively small~\cite{Brandt:2015sxa,Astrakhantsev:2019zkr}. In the case where a continuum extrapolation of correlation function is available, the direct $\chi^2$ fitting using certain ansatze consisting of spectral functions obtained from perturbative NRQCD 
at high temperatures is also used to reconstruct the spectral function~\cite{Ghiglieri:2016tvj,Ding:2016hua,Burnier:2017bod,Ding:2010ga,Ding:2021ise}.

In the past decade, with the great advance of computing hardware and algorithms~\cite{RevModPhys.91.045002,Bedaque:2021bja,Pang:2021vwl,Shanahan:2018vcv,Tanaka:2017niz,Nagai:2020jar,Shi:2021qri,Pang:2016vdc,Yoon:2018krb,Niu:2018csp,Hashimoto:2018bnb,Hashimoto:2018ftp,Zhou:2018ill,Wang:2020tgb,Zhou:2021vza,Carifio:2017bov,Zhao:2021yjo,Nicoli:2020njz,Boyda:2020nfh,Albergo:2021vyo,Ren:2021prq} machine learning has been applied to various topics in physical sciences. Very recently it has been introduced to explore the inverse problem to extract the spectral functions from imaginary Green's function in condensed matter physics~\cite{Yoon:2018,Fournier:2020} and from Euclidean correlators in lattice QCD in high energy nuclear physics~\cite{Kades:2019wtd,Offler:2019eij,Zhou:2021bvw,Horak:2021syv}.
Concerning the type of learning, supervised learning was used in Ref.~\cite{Yoon:2018,Kades:2019wtd,Fournier:2020}, where both Ref.~\cite{Yoon:2018} and Ref.~\cite{Kades:2019wtd} adopted convolutional neural network and Ref.~\cite{Fournier:2020} used fully connected feed-forward layers. On the other hand, machine learning based on the kernel ridge regression, radial basis functions and Gaussian process regression was adopted in Ref.~\cite{Offler:2019eij},  Ref.~\cite{Zhou:2021bvw} and Ref.~\cite{Horak:2021syv}, respectively. Due to the different physics problems the integral kernels (cf. Eq.~\eqref{eq:kernel} in our study) used in the above mentioned studies in machine learning differ from each other, and it is obvious that the integral kernel itself could affect the difficulty level in the reconstruction of the spectral functions. In Refs.~\cite{Kades:2019wtd,Horak:2021syv} the main focus is on the spectral functions related to the Euclidean propagator with the K\"allen-Lehmann spectral representation, i.e. $K(p,\omega)\propto\omega/(\omega^2+p^2)$, and in Ref.~\cite{Offler:2019eij} the spectral function is connected to the Euclidean correlator obtained from lattice QCD in the non-relativistic limit with $K(\tau,\omega)=e^{-\omega\tau}$. In Ref.~\cite{Zhou:2021bvw} a same integral kernel as in our study was adopted, and two types of training spectral functions were used, i.e. training spectral function consisting of 1) two resonance peaks only and 2) one continuum term plus one transport peak. It is also worthy to note that in the most of these mentioned studies the training spectral functions have the same form of the expected reconstructed spectral functions~\cite{Yoon:2018,Offler:2019eij,Kades:2019wtd,Zhou:2021bvw}. Among these studies Refs.~\cite{Kades:2019wtd,Zhou:2021bvw} are restricted to the mock data tests while results of spectral functions reconstructed from lattice QCD correlators are shown in Refs.~\cite{Offler:2019eij,Horak:2021syv}.

 In this paper we present a novel unsupervised neural network SVAE~\footnote{Here we use the term neural network in an extended way, including the network training (e.g. the loss function, etc.).}, based on the variational autoencoder (VAE). The VAE is a generative neural network based on the likelihood criterion, and it is a fertile framework to build new models in and has been used/extended in many areas~\cite{Kingma:2019introVAE}. The VAE is based on the Bayesian theorem and includes two parts: one is the encoder or recognition model and the other is the decoder or generative model. These two support each other and are connected via the distribution over the latent space $\mathcal{Z}$. The former encodes the information of the input data $x$ into a latent space $\mathcal{Z}$ and passes a conditional probability distribution $Q(z|x)$ into a bottleneck architecture, while the latter decodes the latent space $\mathcal{Z}$ that was encoded in the bottleneck layer by the encoder to regenerate the conditional probability $P(x|z)$ of the inputs for given $z$. 
These results from the VAE in the form of the loss function are used to update all parameters within the neural networks, during which the evidence lower bound consisting of a marginal likelihood of the input data and the closeness to the true posterior probability of $z$ given data $x$ is maximized.~\footnote{In the following the subscript $gt$ stands for the ``ground truth value",  referring to information that is known to be real or true, provided by direct observation and measurement as opposed to information provided by inference.} 
 
 Thus the VAE learns to reproduce its input by mapping the input to a latent space in an unsupervised way and it can naturally be used for studying inverse problems. The newly constructed neural network proposed in the current paper contains two setups: the training process consists of two encoders and one decoder, while the reconstruction is made with one encoder and one decoder. The ground truth values of the spectral function and the Euclidean two-point correlator are introduced to adjust the posterior probability distribution via the Kullback-Leibler (KL) divergence and the prior probability distribution via the Shannon-Jaynes entropy~\footnote{The introduction of the entropy term into the neural network gives the name ``SVAE".}.
During the training we adopt training spectral functions of a general form from the Gaussian mixture model which is similar to that in Ref.~\cite{Fournier:2020}, and in the mock data test we test four different kinds of physics motivated hadron spectral function.

The paper is organized as follows. In Section~\ref{sec:setup}, we introduce the framework of neural network . In  Subsection~\ref{subsec:loss} we construct the loss function for the neural network based on the Bayesian theorem, in Subsection~\ref{subsec:paraloss} we present the explicit form of the ingredients of the loss function including the likelihood function and the entropy term, in Subsection~\ref{subsec:topology} we show the topology of the neural network for training and testing, and describe the approach to obtain the final output spectral function and its statistical uncertainties. In Subsection~\ref{subsec:strategy} we provide the form of the training spectral function obtained from the Gaussian mixture model and also the four kinds of physics motivated spectral function for tests. In Subsection~\ref{subsec:mock} we discuss the reconstructed spectral functions in the mock data tests and compare them with those obtained from the MEM. In Section~\ref{sec:lqcd} we apply the SVAE to reconstructing the spectral function from the Euclidean correlator in the pseudoscalar channel obtained in the quenched lattice QCD at one below and the other above the critical temperature, and finally summarize our results in Section~\ref{sec:con}.
Studies on the dependences of reconstructed spectral functions on the noise levels and noise models of mock correlator data are included in  Appendix~\ref{app:dep} and~\ref{app:noise_model}, respectively. In Appendix~\ref{app:inequality} the H\"older's inequality is described for the derivation of Eq.~\ref{eq:evidence}.

\section{Setup of the neural network}
\label{sec:setup}

In this section we introduce our proposed neural network approach, SVAE.
In the following, we suppress the arguments $\tau$ and $T$ of the temporal correlator $G$ and $\omega$ and $T$ of the spectral function $\rho$ to avoid clutter. Thus $\rho$ and $G$ hereafter stand for a vector in $\omega$ and $\tau$ spaces, respectively.

\subsection{Variational autoencoder and the loss function}
\label{subsec:loss}

Based on the Bayes' theorem, the probability of $\rho$ given correlator data $G$ can be expressed as follows
\begin{equation}
P(\rho|G)  = \int dz ~ P(\rho|z, G)P(z|G)\,.
\label{eq:PrhoG}
\end{equation}
Thus, maximizing $P(\rho|G) $ gives the most probable image of $\rho$. Note that the $G$ here can be either averaged correlator over configurations or a sample of correlators obtained on many configurations. Hereafter we denote $G$ for the latter case. The parameter $z$ can be considered as the latent variable in a certain neural network. $P(z|G)$ gives the prior probability distribution of $z$ given data $G$, and $P(\rho|z, G)$ is the posterior probability of $\rho$ given data $G$ with the latent variable $z$ sampled from $P(z|G)$. 
Obviously, this calculation process is compatible with the structure of variational autoencoder (VAE) in machine learning~\cite{kingma2014autoencoding,rezende2014stochastic}. As an unsupervised learning model, VAE is built on the encoder and decoder. For training, the encoder learns to produce a distribution of latent variables fed into the decoder, while the decoder reconstructs the input data; for reconstruction, only the decoder is used as a generative model. In the language of VAE, $P(z|G)$ is generated by an encoder, i.e. mapping the correlator data G into the latent space, while $P(\rho|z, G)$ can be used to construct a decoder to decipher the latent code/variable to extract $\rho$.

In general, the final solution of $\rho$ based on the neural network is obtained by minimizing the loss function. To choose an appropriate loss function we start to work with the logarithm of Eq.~\eqref{eq:PrhoG}, 
\begin{align}
\begin{split}
	\log P(\rho|G) &=\log \bigg(\int \mathrm{d}z ~ P(\rho|z, G)P(z|G) \bigg)\\
&\equiv\log \left(E_{Q(z|G_{gt},\rho_{gt})}\left( P(\rho|z,G) \,\frac{P(z|G)}{Q(z|G_{gt},\rho_{gt})}\right) \right)\\
&\geq E_{Q(z|G_{gt},\rho_{gt})}\left(\log \left(P(\rho|z,G) \,\frac{P(z|G)}{Q(z|G_{gt},\rho_{gt})}\right)\right) \,,
\end{split}
\label{eq:logP1}
\end{align}
where $E_{Q(z|G_{gt},\rho_{gt})} (f(x)) = \int\mathrm{d}z f(x) Q(z|G_{gt},\rho_{gt})$, and the probability distribution $Q(z|G_{gt},\rho_{gt})$, known as the ``ground-truth factor of variations'', is introduced to pick up the ``relevant'' part of $z$ space for $G_{gt}$ and $\rho_{gt}$. Here $G_{gt}$ stands for correlators obtained from the ground truth value of $\rho$ denoted by $\rho_{gt}$ via Eq.~\eqref{eq:Grho}. The last line in the above equation follows from the Jensen's inequality and the concave nature of the logarithm function.

By introducing the Kullback-Leibler (KL) divergence between $Q(z|G_{gt},\rho_{gt})$ and $P(z|G)$ we can further write Eq.~\eqref{eq:logP1} as follows,
\begin{equation}
\log P(\rho|G) \geq E_{Q(z|G_{gt},\rho_{gt})}\bigg[ \log P(\rho|z, G) \bigg] - KL\bigg(Q(z|G_{gt},\rho_{gt})\| P(z|G) \bigg).
\label{eq:logP2}
\end{equation}
The first term on the right hand side of the inequality in Eq.\eqref{eq:logP2} represents the probability of decoding $\rho$ from a latent variable $z$ sampled according to the distribution $Q(z|G_{gt},\rho_{gt})$. The second term in the above equation is the KL divergence, which denotes the ``distance'' between the conditional probabilitiy distributions $Q(z|G_{gt},\rho_{gt})$ and $P(z|G)$ having the following form
\begin{equation}
KL\bigg(Q(z|G_{gt},\rho_{gt})\| P(z|G)\bigg) = \int dz~ Q(z|G_{gt},\rho_{gt})\log\bigg( \frac{Q(z|G_{gt},\rho_{gt})}{P(z|G)} \bigg).
\label{eq:KLQP}
\end{equation}

Thus, based on Eq.~\eqref{eq:logP2} the loss function $\mathfrak{L}$ can be defined as,
\begin{equation}
\mathfrak{L} = - E_{Q(z|G_{gt},\rho_{gt})}\bigg[ \log P(\rho|z, G) \bigg] + KL\bigg(Q(z|G_{gt},\rho_{gt})\| P(z|G) \bigg).
\label{eq:loss1}
\end{equation}
The advantage of the neural network method over the iterative method, such as MEM, is that one knows the ground truth value of $\rho$, which can be used as the prior information of $\rho$ in the training of the neural network. The prior information, $P(\rho|z)$, can be naturally included in the loss function Eq.~\eqref{eq:loss1} since
\begin{equation}
P(\rho|z, G) = \frac{P(\rho|z)P(G|\rho, z)}{P(G|z)}.
\label{eq:PrhozG}
\end{equation}
Here $P(G|z)$ is the normalization factor and $P(G|\rho, z)$ is related to the likelihood function. The loss function can thus be further written as,
\begin{equation}
\mathfrak{L} = - E_{Q(z|G_{gt},\rho_{gt})}\bigg[ \log \frac{P(\rho|z)P(G|\rho, z)}{P(G|z)}\bigg] + KL\bigg(Q(z|\rho_{gt},G_{gt})\| P(z|G) \bigg).
\label{eq:lossfunction}
\end{equation}	  
The aim of the training process is to minimize the loss function, i.e. to maximize the $P(\rho|z,G)$ and minimize the KL divergence term in the above equation. To maximize the $P(\rho|z,G)$ is to find the most probable image of spectral function $\rho$ based on the Bayesian theorem. This is in analogy to the case in the MEM (see e.g. Eq. 3.3 in Ref.~\cite{Asakawa:2000tr}).

\subsection{SVAE and parametrization of the loss function}
\label{subsec:paraloss}
Now, we express various terms in the loss function (cf. Eq.~\eqref{eq:lossfunction}) in details. The first term in the loss function, $P(\rho|z)$, which encodes the prior information of $\rho$, can be expressed as follows inspired by the MEM~\cite{JARRELL1996133,Asakawa:2000tr},
\begin{align}
P(\rho|z) = \frac{1}{Z_S}e^{S},
\label{eq:prior}
\end{align}
where $S$ is the Shannon-Jaynes entropy with the following form~\cite{JARRELL1996133,Asakawa:2000tr}
\begin{align}
S = \int_{0}^{\infty}\mathrm{d}\omega \left(\rho(z) - \rho_{gt} - \rho(z)\log\left(\frac{\rho(z)}{\rho_{gt}}\right) \right) \to \sum\limits_l^{N_\omega} \left(\rho_l(z) - \rho_{gt,l} - \rho_l(z)\log\left(\frac{\rho_l(z)}{\rho_{gt,l}}\right) \right)\,,
\label{eq:entropy}
\end{align}
and $Z_S\approx (2\pi)^{\frac{N_\omega}{2}}$ is the normalization constant with $N_\omega$ the number of points in the $\omega$ space.

We remark there that Eq.\eqref{eq:entropy} has the same functional form as that in the entropy term appearing in the MEM. However, in the MEM, the default model which includes the prior information is an input parameter for the MEM analyses. Thus to make sense of the output spectral function from the MEM one has to study the dependence of the output spectral functions on the input default models. While in the training process of the SVAE, the prior information is fixed to $\rho_{gt}$ which is the ground-truth value of $\rho$ and the best prior information that can be provided. This is a prominent advantage over iterative methods, e.g. MEM. Thus the SVAE is trained to learn to map the most probable spectral function from the various training data sets during the training process, and then extract the spectral function from the input noisy correlator data without any further input parameters.

The second term in the loss function, $P(G|z, \rho)$, is related to the likelihood function and is constructed to have the following form~\cite{JARRELL1996133,Asakawa:2000tr}
\begin{equation}
P(G|z, \rho)  = \frac{1}{Z_L}e^{- L} \,,
\label{eq:likelihood}
\end{equation}
with the likelihood function 
\begin{equation}
L =\sum\limits_{j=\tau_{min}}^{N_{\tau}/2} L_j =\sum\limits_{j=\tau_{min}}^{N_{\tau}/2} \frac{\Big(\hat{G}_j\big[\rho(z)\big]- G_j\Big)^2}{2\alpha_j^2(z)G_j^2} \,.
\label{C2Eq5}
\end{equation}
Since $G(\tau)$ is symmetric at $\tau=N_\tau/2$ and suffers from the lattice cutoff effects in the short distance when computed in lattice QCD, we normally discard a few points in the short distance and use the data points of $G(\tau)$ with $\tau \in [\tau_{min},N_\tau/2]$ during the training process (cf. the dimension of $G$ in the input layer shown in Table.~\ref{tab:hyperpara}). Here both $\rho(z)$ and $\alpha(z)$ are the outputs from the decoder, and $\hat{G}\big[\rho(z)\big]$ is the correlator reconstructed with $\rho(z)$.

We remark here that the $\alpha$ parameter shown in the likelihood function actually has two peculiar features: one is that it can be interpreted as a learned noise level in the likelihood function, which can be different at different distance~\footnote{The lately designed noise model [cf. Eq.~\ref{eq:noise_level}] is consistent with the ignorance of the off-diagonal part in the likelihood function. To put a covariance matrix into the likelihood function one might encounter with numerical issues.  As in our training process, we randomly sample $\rho_{gt}$ and $G_{gt}$, and the chance of having an exact $\rho_{gt}$ is tiny. To have a covariance matrix in the training, we thus need to sample many times $G_{gt}$ produced from a same $\rho_{gt}$. This will increase the numerical cost significantly. From our current results, the current form of likelihood function seems serve the purpose. In the future it would be interesting to study with a covariance matrix in the likelihood function with more computing resources available.}, the other is that it is actually the learned relative weight between the likelihood function and the entropy term. 
We stress here for the latter, since the prominent feature of the SVAE is that it is trained to find the most probable image of the spectral function given noisy correlator data, the key is that the relative weight, $\alpha$, is learned via the training. Note that in the case of MEM, the relative weight is put by hand and the final output spectral function is a weighted sum of spectral functions obtained at different values of $\alpha$~\cite{JARRELL1996133}.

$Z_L$ is a normalization constant with the following form
\begin{equation}
Z_L = \displaystyle \prod_{j=\tau_{min}}^{N_{\tau}/2} \sqrt{2\pi\alpha_j^2(z)G_j^2} \,.
\label{C2Eq6}
\end{equation}
The evidence or the normalization factor $P(G|z)$ can be expressed using Eq.~\eqref{eq:prior} and Eq.~\eqref{eq:likelihood} as follows
\begin{equation}
\begin{aligned}
P(G|z) &= \int\!\! \mathcal{D}\rho P(G|z, \rho)P(\rho|z)
\quad      = \int\!\! \mathcal{D}\rho\ \frac{1}{Z_SZ_L}e^{-L+S} \\
&\leq \frac{\Lambda^{\frac{1}{p}}}{Z_LZ_S}\prod\limits_{j=\tau_{min}}^{N_{\tau}/2} \left[\frac{2\pi\alpha_j^2(z)G_j^2}{p}\right]^{\frac{1}{2p}}\left[\frac{2\pi}{q}\right]^{\frac{N_\omega}{2q}}\,,
\end{aligned}
\label{eq:evidence}
\end{equation} 
where $p$ and $q$ are positive numbers with $\frac{1}{p}+\frac{1}{q}=1$ and $p,q>1$, while $\Lambda$ is a constant depending only on the integral kernel $K(\tau, \omega, T)$ (cf. Eq.~\eqref{eq:kernel}). The derivation of the inequality is based on the $H\ddot{o}lder$ inequality and can be found in Appendix~\ref{app:inequality}.

For the conditional probability distributions $Q(z|\rho_{gt}, G_{gt})$ and $P(z|G)$ in the loss function, we parameterize them with the Gaussian mixture model~\cite{bishop2006pattern} as follows
\begin{align}
\label{eq:Qpara}
Q(z|\rho_{gt},G_{gt})&=\sum\limits_{i}^{N_g} \pi_{i}\prod\limits_{k}^{N_z} \mathcal{N}(\mu_{i,k}, \sigma_{i,k}), \\
P(z|G) &=\sum\limits_{i}^{N_g} \hat{\pi}_{i}\prod\limits_{k}^{N_z} \mathcal{N}(\hat{\mu}_{i,k}, \hat{\sigma}_{i,k}).
\label{eq:Ppara}
\end{align}
Here $\mathcal{N}(\mu_{i,k}, \sigma_{i,k})$ ($\mathcal{N}(\hat{\mu}_{i,k}, \hat{\sigma}_{i,k})$) are Gaussian distributions of $z$ with the mean value $\mu_{i,k}$ ($\hat{\mu}_{i,k}$) and the variance $\sigma_{i,k}$ ($\hat{\sigma}_{i,k}$). $k$ ranges from 1 to $N_z$ with $N_z$ being the dimension of the hidden space $\mathcal{Z}$. $\pi_{i}$ and $\hat{\pi}_i$ are the weights (probabilities) for each Gaussian component with $\sum_{i}^{N_g} \pi_i=1$ and $\sum_{i}^{N_g} \hat{\pi}_i=1$. $N_g$ is the number of normal distributions $\mathcal{N}$ used in the Gaussian mixture model. $\pi_i$ ($\hat{\pi}_i$),  $\mu_{i,k}$ ($\hat{\mu}_{i,k}$), and $\sigma_{i,k}$ ($\hat{\sigma}_{i,k}$) are outputs of the encoders (cf. Fig.~\ref{fig:endcoder}) and are used to sample the latent variable $z$ according to the distributions Eqs.~(\ref{eq:Qpara}-\ref{eq:Ppara}).

Based on the parameterization in Eq.~\eqref{eq:Qpara} and~\eqref{eq:Ppara} the upper boundary of KL divergence in Eq.~\eqref{eq:lossfunction} can thus be expressed as follows~\cite{Hershey2007ApproximatingTK}
\begin{equation}
KL\left(Q(z|\rho_{gt}, G_{gt})\| P(z|G)\right) \leq \sum\limits_{i=1}^{N_g} \pi_i\Big [ \ KL\Big(\mathcal{N}(\mu_{i}, \sigma_{i})\|\mathcal{N}(\hat{\mu}_{i}, \hat{\sigma}_{i})\Big) +\log\Big(\pi_i /\hat{\pi}_i\Big) \Big] \,.
\label{eq:upperboundKL}
\end{equation}
With the choice of $q=2\pi$ in Eq.~\eqref{eq:evidence} and neglecting unnecessary constants, e.g. $Z_S$,  combining Eqs.~\eqref{eq:lossfunction}, \eqref{eq:prior}, \eqref{eq:likelihood}, \eqref{eq:evidence}, \eqref{eq:upperboundKL} the loss function in the SVAE can be finally written as follows,
\begin{align}
\mathfrak{L}_{\mathrm{SVAE}}=& \hspace*{-7mm} \int\limits_{z\sim Q(z|\rho_{gt}, G_{gt})} \hspace*{-4mm} \Bigg\{
 \sum\limits_l^{N_\omega} \left[\rho_{gt,l}-\rho_l(z) +\rho_l(z)\log\left(\frac{\rho_l(z)}{\rho_{gt,l}}\right)\right] +\sum\limits_{j=\tau_{min}}^{N_\tau/2 
 } \frac{1}{2}\left[ \bigg( \frac{\hat{G}_j\big[\rho(z)\big]-G_j}{\alpha_j(z)G_j}\bigg)^2 + \frac{2\pi-1}{2\pi}\log\left(\alpha^2_j(z)G_j^2\right)\right] \Bigg\} dz \nonumber \\
+&  \sum\limits_{i=1}^{N_g} \Bigg[\frac{\pi_i}{2} \sum_{k=1}^{N_z} \Bigg(\frac{\left(\mu_{i,k} - \hat{\mu}_{i,k}\right)^2}{\hat{\sigma}_{i,k}^{2}} + \mathrm{log}\frac{\hat{\sigma}_{i,k}^{2} }{\sigma_{i,k}^2} +  \frac{\sigma_{i,k}^2}{\hat{\sigma}_{i,k}^{2}} - 1\Bigg) + \pi_i \log\Big(\pi_i /\hat{\pi}_i\Big)\Bigg] \,.
\label{eq:lossbound}   
\end{align}	
In the above equation an entropy term, i.e. Eq.~\eqref{eq:entropy} is used (the first term in square brackets) and this brings one of the prominent differences of our proposed neural network approach from others. We thus call the neural network using this loss function SVAE.

We remark here that the loss function adopted in our study is different from the commonly used ones, e.g. mean absolute value of spectral function (cf. Eq. 10 in Ref.~\cite{Fournier:2020}) and also correlators (cf. Eq. 10 in Ref.~\cite{Kades:2019wtd}). The usage of the entropy term and the likelihood function in the loss function is inspired by the MEM (cf. Eq. 3.3 in Ref.~\cite{Asakawa:2000tr}). According to the loss function (cf. Eqs.~\ref{eq:lossfunction} and~\ref{eq:lossbound}) the SVAE is thus trained to obtain the most probable image of the spectral function as the number of solutions to the inverse problem by default is infinite. Note also that $\alpha(z)$, which controls the relative strength between the likelihood and the entropy, is an output from the ``Decoder" (cf. Fig.~\ref{fig:endcoder}) during the training.

It is also noteworthy to compare the SVAE to the conditional VAE (CVAE)~\cite{kingma2014autoencoding}. Although the CVAE and SVAE appear to be similar in the sense that the generative process is conditioned on an input, there exists a key difference between CVAE and SVAE which makes them applicable to different problems. The CVAE only allows us to tackle problems where the input-to-output mapping is one-to-many~\cite{kingma2014autoencoding}, while the SVAE allows us to deal with problems where the input-to-output mapping is many-to-many, since the correlator data as the input in our case is stochastic and not fixed, i.e. may get values not presented during training. This difference originates from the provided $G_{gt}$ and $\rho_{gt}$ as well as the entropy term in the loss function.

\subsection{Topology of the SVAE}\label{section3}
\label{subsec:topology}
\begin{figure}[!thp]
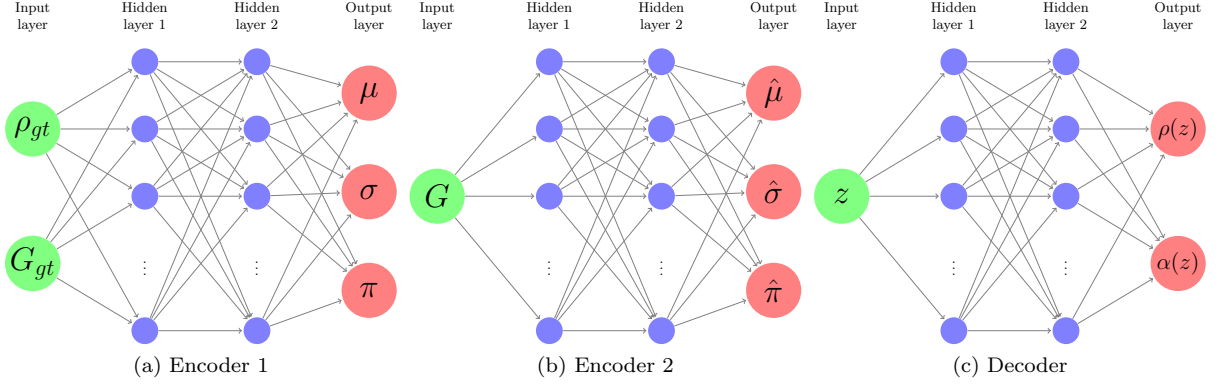

	\begin{center}
		\subfloat[Encoder 1]{\includegraphics[]{Encoder1.tikz}\label{Fig1(a)}}
		\subfloat[Encoder 2]{\includegraphics[]{Encoder2.tikz}\label{Fig1(b)}}
		\subfloat[Decoder]{\includegraphics[]{Decoder.tikz}\label{Fig1(c)}}
	\end{center}
	\caption{Left: The encoder used to convert the ground truth values, $\rho_{gt}$ and $G_{gt}$, into parameters, $\mu$, $\sigma$ and $\pi$ describing the $z$ distribution in the Gaussian mixture model. This encoder maps a $\rho_{gt}$ - $G_{gt}$ pair to the parameters of the probability distribution of the latent variable $z$, $Q(z|\rho_{gt},G_{gt})$ via Eq.~(\ref{eq:Qpara}). Middle: Similar as the left plot but for the input being the stochastic data, $G$. This encoder produces the probability distribution $P(z|G)$ through Eq.~(\ref{eq:Ppara}). Right: The decoder used to convert the latent variable $z$ to the output spectral function $\rho(z)$ and $\alpha(z)$. In all the three figures, the green dots stand for the inputs while the red dots denote the outputs from (de)(en)coders. }
	\label{fig:endcoder}
\end{figure}

The design of topology of the neural network, SVAE, is based on Eq.~\eqref{eq:lossfunction} and Eq.~\eqref{eq:lossbound}. The basic components are the decoder and encoder. ``Encoder 1" shown in the left plot of Fig.~\ref{fig:endcoder} encodes the information of the ground truth values, $\rho_{gt}$ and $G_{gt}$, and their correlations, into the latent space. This encoder thus results the distribution $Q(z|\rho_{gt},G_{gt})$. On the other hand, ``Encoder 2" shown in the middle plot of Fig.~\ref{fig:endcoder} encodes the information of the input (training) stochastic data $G$~\footnote{The construction of $G$ for the mock data is described in Section~\ref{subsec:strategy}.} into the latent space, and thus related to the distribution $P(z|G)$. Both probability distributions $Q(z|\rho_{gt},G_{gt})$ and $P(z|G)$ are parameterized by the outputs of the encoders, $\mu~(\hat{\mu})$, $\sigma ~(\hat{\sigma})$ and $\pi~(\hat{\pi})$, i.e., the mean, variance and weight values of the Gaussian distribution $\mathcal{N}$ in the Gaussian mixture model. The ``Decoder", shown in the right plot of Fig.~\ref{fig:endcoder}, is fed with a vector $z$ sampled from the distributions $Q(z|\rho_{gt},G_{gt})$ or $P(z|G)$. It then produces the prediction for the spectral function, $\rho(z)$ on the output.

\begin{table}[htbp]
	\centering
	\begin{tabular}{|c|c|c|c|c|c|}
		\hline
		\diagbox{\scriptsize{Subnet}}{\scriptsize{Layer}}& \scriptsize{Input layer}& \scriptsize{Hidden Layer 1} &\scriptsize{Hidden Layer 2}& \scriptsize{Output layer}\\
		\hline
		\multirow{3}{*}{\scriptsize{Encoder 1}} &\multirow{3}{*}{$\rho_{gt}\& G_{gt}: N_\omega+N_{\tau}/2-\tau_{min}+1$}&\multirow{3}{*}{500, Relu} & \multirow{3}{*}{250,~Hard sigmoid} &\multicolumn{1}{c|}{$\mu : N_g \times N_z $, none} \\
		\cline{5-5}
		& & & &$\sigma: N_g \times N_z$, softplus \\ 
		\cline{5-5}
		& & & &$\pi: N_g$, softmax \\ 
		\hline
		
		\multirow{3}{*}{\scriptsize{Encoder 2}}& \multirow{3}{*}{$G:N_{\tau}/2-\tau_{min}+1$} & \multirow{3}{*}{100,~Relu}  & \multirow{3}{*}{250,~Hard sigmoid}& \multicolumn{1}{c|}{$\hat{\mu}: N_g\times N_z$, none}\\
		\cline{5-5}
		& & & &$\hat{\sigma}: N_g \times N_z$,~softplus \\
		\cline{5-5}
		& & & &$\hat{\pi}: N_g$,~softmax \\
		\hline
		
		\multirow{2}{*}{\scriptsize{Decoder}}&\multirow{2}{*}{$z:N_z$}& \multirow{2}{*}{250,~Relu} & \multirow{2}{*}{500,~Hard sigmoid}  &\multicolumn{1}{c|}{$\rho(z):  N_\omega$, softplus} \\
		\cline{5-5}
		& & & &$\alpha(z): N_{\tau}/2-\tau_{min}+1$,~softplus 
		\\ \hline
	\end{tabular}
	\caption{The dimension of $\rho_{gt}$, $G_{gt}$, $G$ and $z$ in the input layer and $\mu_j$ ($\hat{\mu}_j$), $\sigma_j$ ($\hat{\sigma}_j$), $\pi_j$ ($\hat{\pi}_j$), $\rho(z)$ and $\alpha(z)$ in the output layer, the number of neurons and type of activation function used in the neural network as shown in Fig.~\ref{fig:endcoder}. In our current study $N_\omega$=10000, $\tau_{min}$=4, $N_g$=5 and $N_z$=50 are adopted.}
	\label{tab:hyperpara}
\end{table}

All the networks shown Fig.~\ref{fig:endcoder} are built with similar structures, i.e. each of them has one input layer, two hidden layers and one output layer. The dimension of parameters, the number of neurons and the types of activation functions for each layer of the network are listed in Table~\ref{tab:hyperpara}. The input layer is fed with $\rho_{gt}$ and $G_{gt}$ in ``Encoder 1", $G(\tau)$ in ``Encoder 2" and $z$ in the ``Decoder", each of which is considered as a vector and has a dimension of $N_\omega+N_\tau/2-\tau_{min}+1$, $N_\tau/2-\tau_{min}+1$ and $N_z$, respectively.
To avoid the consumption of large memories as well as the oscillation of the loss function, we use the batch normalization~\cite{ioffe2015batch} on the input and two hidden layers in all the neural networks. The training was done with a mini-batch size of 100, and typically five million epochs were used for TensorFlow version 15.0 on NVIDIA V100 GPUs. Concerning the activation function we use the rectified linear function (Relu) for the first hidden layer and a hard sigmoid in the second hidden layer. In the ``Hidden layer 1" the number of neurons adopted in our study are 500, 150 and 250 for ``Encoder 1", ``Encoder 2" and ``Decoder", respectively, while in the ``Hidden layer 2" they are 250, 250 and 500. On the output layer of ``Encoder 1" and ``Encoder 2" we use the softplus and softmax function for $\sigma$ and  $\pi$, respectively. For the output layer of the ``Decoder" we use the softplus activation function for both $\rho$ and $\alpha$. The dimensions of $\hat{\pi}~(\pi)$, $\hat{\mu}~(\mu)$, $\hat{\sigma}~(\sigma)$, $\rho$ and $\alpha$ in the output layer are $N_g$, $N_g\times N_z$, $N_g\times N_z$, $N_\omega$ and $N_\tau/2-\tau_{min}+1$, respectively.

Finally, we use the Adam optimizer~\cite{kingma2014adam}, which assigns different learning rates to each parameter to speed up ability of the neural network to optimize itself. In addition, this method has a very good robustness and is widely used in various non-convex optimization problems in machine learning.

\begin{figure}[!thpb]
	\centering
	\includegraphics[]{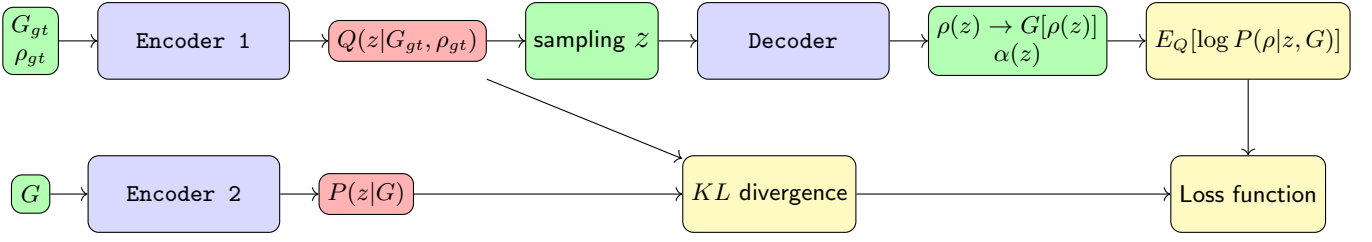}
	\caption{Training process involving the encoders and decoder shown in Fig.~\ref{fig:endcoder}.
	}
	\label{fig:training}
\end{figure}

\begin{figure}[!thpb]
	\centering
	\includegraphics[]{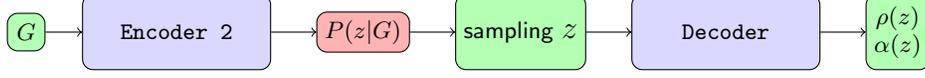}
	\caption{Reconstruction process.}
	\label{fig:reconstruction}
\end{figure}
The building blocks of the designed neural network, i.e. the encoders and decoder shown in Fig.~\ref{fig:endcoder} are used in the training process to minimize the loss function. During training $\rho_{gt}$ and $G_{gt}$ as well as $G$ are firstly fed simultaneously into the ``Encoder 1" and ``Encoder 2", respectively (see Fig.~\ref{fig:training}). The conditional probability distributions of $z$, $Q(z|G_{gt},\rho_{gt})$ and $P(z|G)$, can then be obtained from these two encoders. With these two probability distributions the KL divergence, KL$(Q(z|G_{gt},\rho_{gt})||P(z|G))$, which is one of the two ingredients of the loss function (cf. Eq.~\eqref{eq:lossfunction}), can be obtained. The other ingredient of the loss function, $E_{Q(z|G_{gt},\rho_{gt})}[\log(P(\rho|z,G))]$, can be obtained by using $\rho(z)$. Here $\rho(z)$ is produced from the ``Decoder" fed by $z$ as sampled from $Q(z|G_{gt},\rho_{gt})$. By minimizing the loss function (cf. Eq.~\eqref{eq:lossbound}), the parameters of the neural networks can thus be determined.

Once the training is done we drop ``Encoder 1" and use the setup based on the trained ``Encoder 2" and the ``Decoder" to extract the spectral function  $\rho(z)$ from the stochastic correlator data $G$ with $N_c$ configurations (see  Fig.~\ref{fig:reconstruction}). 
Since the procedure involves a sampling of the latent variable $z$, we need to average over the latent space at a certain epoch $\Theta$. Furthermore, we also need to average over $N_c$ configurations. 
To obtain the spectral function $\tilde{\rho}_{\Theta}$ we start from Eq.~(\ref{eq:PrhoG}) and have

\begin{equation}
\begin{aligned}
\tilde{\rho}_{\Theta}\equiv\frac{1}{N_c}\sum\limits_{m=1}^{N_c}\tilde{\rho}_{\Theta,m}  &=\frac{1}{N_c}\sum\limits_{m=1}^{N_c}\int \mathrm{d}\rho ~\rho P(\rho |G_m) \\
&=\frac{1}{N_c}\sum\limits_{m=1}^{N_c}\int \mathrm{d}\rho \int \mathrm{d}z ~\rho ~ P(\rho|z, G_m)P(z|G_m) \\ 
&=\frac{1}{N_c}\sum\limits_{m=1}^{N_c}\int \mathrm{d}\rho \int \mathrm{d}z ~\rho  ~ \frac{P(\rho|z)P(G_m|\rho, z)}{P(G_m|z)}~P(z|G_m)\,.
\end{aligned}
\label{eq:der_rho1}
\end{equation}
 Here $G_m$ stands for the $m$-th configuration of input correlator data, and $N_c$ is the total number of configurations having the order of $\mathcal{O}(100)$.
Since $P(G|z) \propto \left(\alpha^2(z)G^2\right)^{\frac{1}{2p}}$ as implemented in the loss function (Eq.~\ref{eq:lossbound}) (cf. the last line of Eq.~\ref{eq:evidence}), and after the training the 'Decoder' gives $P(\rho|z)\propto\delta(\rho(z)-\rho)$, Eq.~\ref{eq:der_rho1} then becomes
\begin{equation}
\begin{aligned}
\tilde{\rho}_{\Theta}   &= \frac{1}{N_c}\sum\limits_{m=1}^{N_c}\int_{z\sim P(z|G)} \mathrm{d}z ~\rho(z) ~\exp\left(-\sum_{j=\tau_{min}}^{N_\tau/2}\frac{\left(\hat{G}_{m,j}[\rho(z)]-G_{m,j}\right)^2}{2\alpha_j^2(z)G_{m,j}^2}\right)
\left(\prod_{j=\tau_{min}}^{N_\tau/2}\big{(}\alpha_j^2(z)G_{m,j}^2\big{)}^{-\frac{1}{2p}}\right)  \\
&\to \frac{1}{N_cN_s}\sum\limits_{m=1}^{N_c}  \sum\limits_{n=1}^{N_s}  \rho(z_n) ~\exp\left(-\sum_{j=\tau_{min}}^{N_\tau/2}\frac{\left(\hat{G}_{m,j}[\rho(z_n)]-G_{m,j}\right)^2}{2\alpha_j^2(z_n)G_{m,j}^2}\right) \left(\prod_{j=\tau_{min}}^{N_\tau/2}\big{(}\alpha_j^2(z_n)G_{m,j}^2\big{)}^{-\frac{1}{2p}}\right)\,.
\end{aligned}
\label{eq:rho_over_z}
\end{equation}
Note that the trivial constants, e.g. $Z_L$, $Z_S$ etc, drop out in the evaluation of $P(\rho|z,G_m)$. The second line in the above equation is the summed version used in practice and $z_n$ sampled from $P(z_n|G_m)$  represents 
the $n$-th sample with $N_s$ having the order of $\mathcal{O}(1000)$. We remark here that all the quantities shown in the above equation are actually computed on each training epoch, where the parameters of the neural network, e.g. the network weights etc denoted by $\Theta$, are obtained. However, due to the feature of mini-batch training strategy and feedback mechanism, the resulting value of the loss function with $\Theta$ fluctuates around its true minimum value with $\Theta^*$, leading to the fluctuation of these obtained parameters around the local solution $\Theta^*$. Thus to take into account the variation of $\Theta$ from $\Theta^*$, we use the following bootstrap method to compute the final output spectral function $\overline{\rho}$ and its uncertainty.
. 

 In the bootstrap method we draw $N_{btp}$ bootstrap samples to compute the error of our final output spectral functions. In each bootstrap sample we firstly randomly choose an epoch ${\Theta}$ (uniformly), and then at this certain epoch we sample $N_c$ times of $\tilde{\rho}_{{\Theta},m}$ uniformly. Note the repetition of $\tilde{\rho}_{{\Theta},m}$ and $\Theta$ in the bootstrap samples is allowed. 
The final mean value $\overline{\rho}$ and its uncertainty $\sigma_{\rho}$ are thus expressed as follows: 
\begin{align}
 \overline{\rho} & = \frac{1}{N_{btp}} \sum_{i=1}^{N_{btp}} \tilde{\tilde{{\rho}}}_{i}\,, \\
    \sigma_{\rho} & = \sqrt{\frac{1}{N_{btp}}\sum_{i=1}^{N_{btp}}(\tilde{\tilde{{\rho}}}_{i}-\overline{\rho})^2} \,\label{eq:sigma_rho},
\end{align}
where $\tilde{\tilde{{\rho}}}_{i}$ is the average value of spectral function $\tilde{\rho}_{{\Theta},m}$ over $N_c$ sub-samples in the configuration space at a fixed $\Theta$ in the $i$-th bootstrap sample. Note that $\sigma_{\rho}$ also includes the statistical uncertainties from the input correlator data ${G}$ as which enters into Eq.~\eqref{eq:sigma_rho} implicitly through Eq.~\eqref{eq:rho_over_z}.

\section{Training Strategy and mock data tests}

\subsection{Spectral functions used in the training and for tests}
\label{subsec:strategy}
To train the neural networks we adopt the following form of the spectral function based on the Gaussian mixture model~\cite{bishop2006pattern},
\begin{equation}
\begin{aligned}
\frac{{\rho}_{train}}{{\omega}^2} = \, \hat{\theta}({\omega}, \delta_1, \zeta_1) \bigg( \sum\limits_{i=1}^{\hat{N}_g}C_ie^{-(\frac{{\omega}-M_i}{\gamma})^2} \left(1-\hat{\theta}({\omega}, \delta_2, \zeta_2)\right) +C_{i=0} \hat{\theta}({\omega}, \delta_2, \zeta_2) \bigg)
\,.
\end{aligned}
\label{eq:training_rho}
\end{equation} 
Here $\hat{\theta}(\hat{\omega}, \delta, \zeta)$ is a cut-off function having the following sigmoid form
\begin{equation}
\hat{\theta}({\omega}, \delta_i, \zeta_i) = \frac{1}{1+\exp(-\frac{\omega-\delta_i}{\zeta_i})} \,.
\label{eq:cut} 
\end{equation}
$\hat{\theta}({\omega}, \delta_1, \zeta_1)$ is introduced to suppress the spectral function at $\omega\approx0$ while $\hat{\theta}({\omega}, \delta_2, \zeta_2)$ is used to smoothly connect the resonance peak and the continuum part of the spectral function.
The number of Gaussian peaks, $\hat{N}_g$, is set to $\hat{N}_g=50$, which shall be sufficient to reproduce a general spectral function. In our training all the parameters are dimensionless. In other words, all the quantities are scaled with (inverse) lattice spacing in the lattice QCD simulations. For instance the mass of $\eta_c$, $M_{\eta_c}=2.98$ GeV, corresponds to a dimensionless number of $\simeq2.98/18.97\sim0.15$ in our computation with 18.97 GeV being the inverse lattice spacing in the state-of-the-art quenched lattice simulations~\cite{Ding:2012sp}. 

\begin{figure}[!htbp]
\centering
	\includegraphics[width=0.5\textwidth]{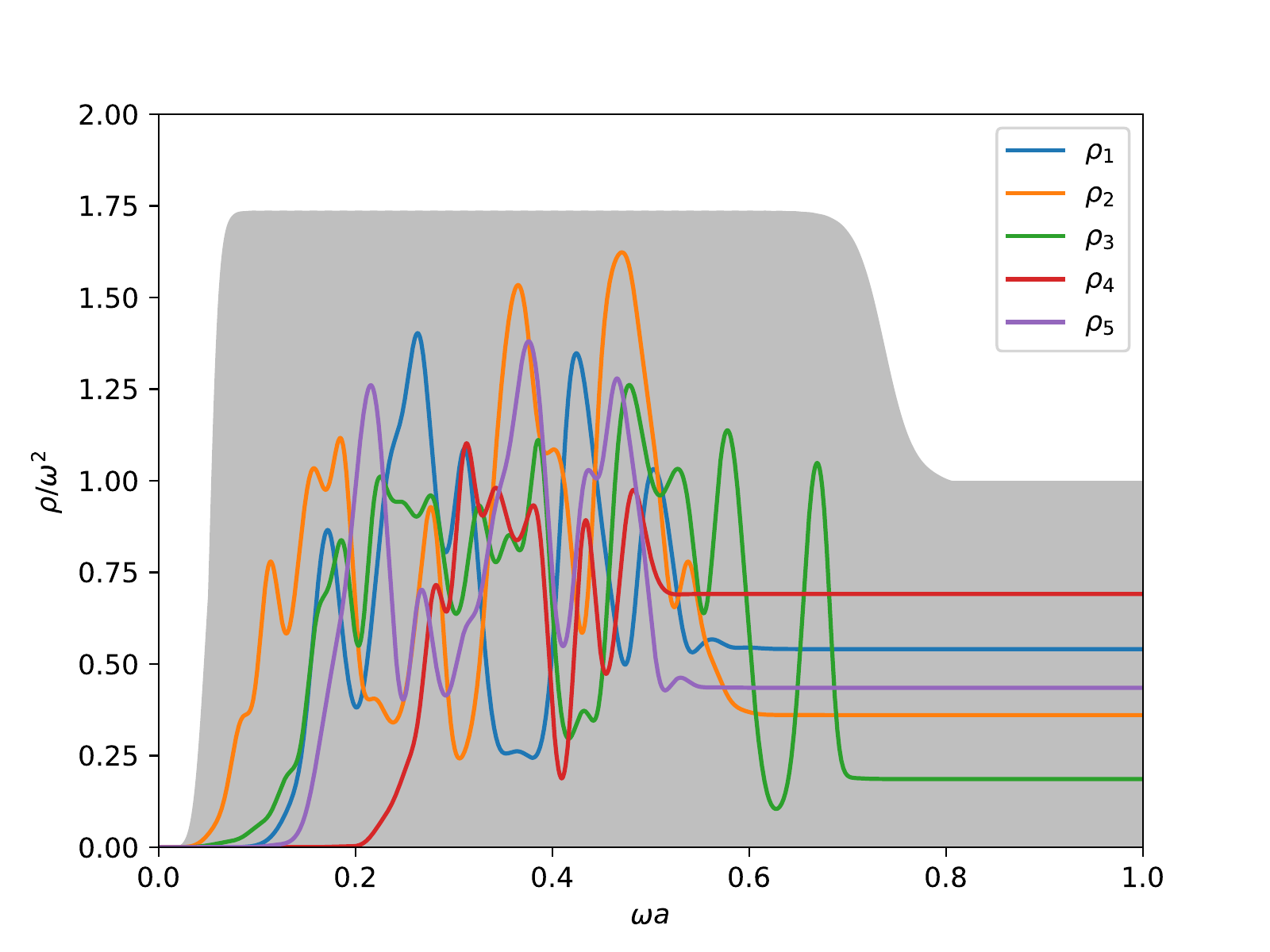}
	\caption{$\rho_i$ with $i={1,2,3,4,5}$ are shown as examples of spectral functions used in the training which are sampled randomly according to Eq.~(\eqref{eq:training_rho}) with parameters listed in Table~\ref{tab:traning_window}. The grey shadow area shows the full sampled range of $\rho_{train}$ according to Eq.~\eqref{eq:training_rho} and Table~\ref{tab:traning_window}.
	}
	\label{fig:training_sample}
\end{figure}

To cover the interesting $\omega$ region the peak location of the Gaussian distribution, $M_i$, is chosen uniformly in the range [0.05, 0.8] with $i$ ranging from 1 to $\hat{N}_g$ and a step length of $(0.8-0.05)/\hat{N}_g$. This corresponds to $M_i$ ranging from $\simeq M_{\eta_c}/3$ to $\simeq 5M_{\eta_c}$. $C_i$ is the amplitude for the $i$-th Gaussian peak and is sampled uniformly in the range of [0, 1].
The value of $\delta_1$, which could be served as the threshold of the spectral function becoming nonzero, is sampled uniformly in the range of [$m_c$, 3$m_c$] with charm quark mass $m_c \sim (1./20, 2./20)$, considering that the threshold is 2$m_c$ in the free limit.
While the value of $\delta_2$, which represents the starting point of a possible continuum part, is assumed to be larger than $\delta_1$ and sampled conditionally in the region of [$\delta_1+0.1$, $\delta_1+0.6$].  $\zeta_1$($\zeta_2$) adjusts the sharpness of the cut-off function, and its value is sampled uniformly from $[0.005, 0.02]$. The detailed information for parameters used in the training spectral function is listed in Table~\ref{tab:traning_window}.
The resulting spectral functions used in the training are also visualized as shown in Fig.~\ref{fig:training_sample}.

\begin{table}[htbp]
	\centering
	\begin{tabular}{|c|c|c|c|}
		\hline
		Parameters & interval  & Parameters & interval  \\ \hline
		$M_ i$ & $[0.05, 0.8]$  &  $\zeta_2(\sim[0.1,0.2]m_c)$& $[0.005, 0.02]$    \ \\ \hline
		$C_i$ & $[0,1]$ & $\delta_1(\sim[1,3]m_c)$& $[0.05, 0.3]$ \\ \hline
		$\zeta_1(\sim[0.1,0.2]m_c)$& $[0.005, {0.02}]$ &$\delta_2(\sim 8m_c)$& $[\delta_1+0.1, \delta_1+0.6]$ \\ \hline
		
	\end{tabular}
	\caption{Adopted parameters in Eq.~\eqref{eq:training_rho} in our study. Additionally we use $N_\omega$=10000, $\tau_{min}$=4, $N_g$=5, $\hat{N}_g=50$ and $N_z$=50 in the training.
	}
	\label{tab:traning_window}
\end{table}

Thus the basic strategy for the training is summarized as follows:
\begin{enumerate}
	\item We sample $\rho_{train}$ randomly according to Eq.~\eqref{eq:training_rho} with the parameters listed in Table~\ref{tab:traning_window}. Then we feed $\rho_{train}$ and resulting temporal correlators according to Eq.~\eqref{eq:Grho} as $\rho_{gt}$ and $G_{gt}$ to the ``Encoder1" (cf. Fig.~\ref{fig:training}).
	
	\item For each $G_{gt}(\tau)$ we add a Gaussian noise constructing the stochastic data $G(\tau)$ to be fed to the ``Encoder 2" (cf. Fig.~\ref{fig:training}). Thus $G(\tau)$ is sampled from a Gaussian distribution with its mean value $G_{gt}$ and its  variance having the following form~\footnote{Here we followed the strategy used in Ref.~\cite{Asakawa:2000tr} and adopted a Gaussian noise to the correlator. We also tried a log-normal noise~\cite{DeGrand:2012ik} and the corresponding results are shown in Appendix~\ref{app:noise_model}.},
	\begin{equation}
b(\tau)\times {G}_{gt}(\tau).
\label{eq:noise_level}
	\end{equation}
	Here $b(\tau)=\sigma_{lat}(\tau)/G_{lat}(\tau)$ and the variance is chosen to mimic the noise level of the lattice QCD data, i.e. 
	$b(\tau)$ becomes larger as $|\tau-N_\tau/2|$ becomes smaller and the largest relative error is $b(N_\tau/2)=1.5\%$.~\footnote{
	The dependence of the reconstruction on the noise level is given in Appendix~\ref{app:dep}}. Note the values of the noise level $b(N_\tau/2)$ in the training data are the same as used in the mock data tests for consistency.
	\item The feeding process in each training epoch described in procedures 1 and 2 is performed simultaneously with one mini-batch containing 100 training samples. 
	\item For each value of $N_\tau$ we repeat the training process described in procedures 1, 2 and 3.
	\end{enumerate}

In our current study we set $N_\omega$=10000, $\tau_{min}$=4, $N_g$=5, $\hat{N}_g=50$ and $N_z$=50, and train the SVAE with two values of $N_\tau$, i.e. 48 and 96.

For the tests of the neural network in SVAE we construct spectral functions consisting of following two different kinds of physics motivated spectral functions. They are listed as follows:
\begin{enumerate}
	\item One resonance peak combined with a continuum spectral function
	\begin{equation}
	\begin{aligned}
	{\rho}_{res+cont} = &\zeta({\omega}, M_{res}, \Gamma){\rho}_{res}({\omega}, C_{res}, M_{res}, \Gamma)(1 - \zeta({\omega}, M_{res}+\Gamma, \Gamma)) \\
	& + \zeta({\omega}, M_{res}+\Gamma, \Gamma) {\rho}_{cont}(C_{cont}, M_{cont}) \,,
	\end{aligned}
	\label{eq:res+cont}
	\end{equation}
where $\zeta(\omega, M_{res},\Delta)=1/(1+e^{\frac{M_{res}^2-\omega^2}{\omega \Delta}})$ is a cut-off function smoothing out the constructed spectral function, and the resonance peak and the continuum part of the spectral function, $\rho_{res}$ and $\rho_{cont}$, are expressed as follows
	\begin{align}
{\rho}_{res}  &= C_{res} \,\frac{{\omega}^2}{(\frac{{\omega}^2}{M_{res}\Gamma} - \frac{M_{res}}{\Gamma})^2  + 1} \,,
\label{eq:res} \\
	{\rho}_{cont} &=C_{cont} \,\frac{3{\omega}^2}{8\pi} \theta({\omega}^2 - 4M_{cont}^2)\,\mathrm{tanh}\Big(\frac{{\omega}}{4T}\Big)\sqrt{1 - \Big(\frac{2M_{cont}}{{\omega}} \Big)^2}\,\Big(2+\Big(\frac{2M_{cont}}{{\omega}}\Big)^2\Big) \,.
\end{align}
Here ${\rho}_{res} $ has a relativistic Breit-Wigner form while $\rho_{cont}$ is the continuum part in the free limit~\cite{Karsch:2003wy,Aarts:2005hg}. $\theta(x)$ is a step function having $\theta(x)=1$ with $x>0$ and $\theta(x)=0$ otherwise.

Based on Eq.~\eqref{eq:res+cont} we will use three different spectral functions for tests, i.e. only one resonance peak, only one continuum part, and one resonance peak with one continuum part.

\item Non-relativistic QCD (NRQCD) motivated perturbative spectral function~\cite{Burnier:2017bod}
\begin{equation}
\label{eq:pertspf}
\rho_{pert}(\omega)= A^{match} \rho^{\text{pNRQCD}}(\omega)\theta(\omega^{match}-\omega)+\rho^{vac}(\omega)\theta(\omega-\omega^{match}).
\end{equation}
Here $A^{match}$ and $\omega^{match}$ are determined coefficients used to reproduce the continuum extrapolated charmonium correlator data in the pseudoscalar channel, and $\rho^{\text{pNRQCD}}(\omega)$ is obtained from the pNRQCD computation while $\rho^{vac}$ is from the perturbative computation in the vacuum asymptotics. The parameters are adopted from Ref.~\cite{Burnier:2017bod}.
\end{enumerate}

The typical time cost in the training process with $N_\tau = 96$ and $48$ data samples is around 8 hours and 6.5 hours for 1 million epochs using one NVIDIA V100 GPU, respectively. After the training is done, the cost for each test is less than 2 minutes.

\subsection{Results from mock data tests}
\label{subsec:mock}
First, we test our network on mock data generated using Eq.~\eqref{eq:res+cont}. In each test we calculated the ground truth $G_{gt}$ from $\rho_{gt}$ and generated 200 configurations of $G$ to be fed to the network. To be consistent we use the the same strategy adopted in the training process, cf. Section III A and Eq.~\ref{eq:noise_level} to produce noisy level of the mock data. Here the same value of $b(\tau)$ is the same as those adopted in the training process.
The common criteria on the reconstruction quality of the spectral function is the shift of the peak location and the change of the peak height.~\footnote{In the case of free continuum spectral function where there are no peaks, the output spectral function is thus considered to be bad or incorrect if fake peaks appear.} As we will see later, in most cases the peak location can be better reproduced compared to the peak height. It is also noteworthy to mention that the correlators computed from the output spectral function in either the MEM~\cite{Ding:2017std} or SVAE analyses always agree with the input correlator data within errors.

\begin{figure}[!htp]
	\centering
	\includegraphics[width=0.45\textwidth]{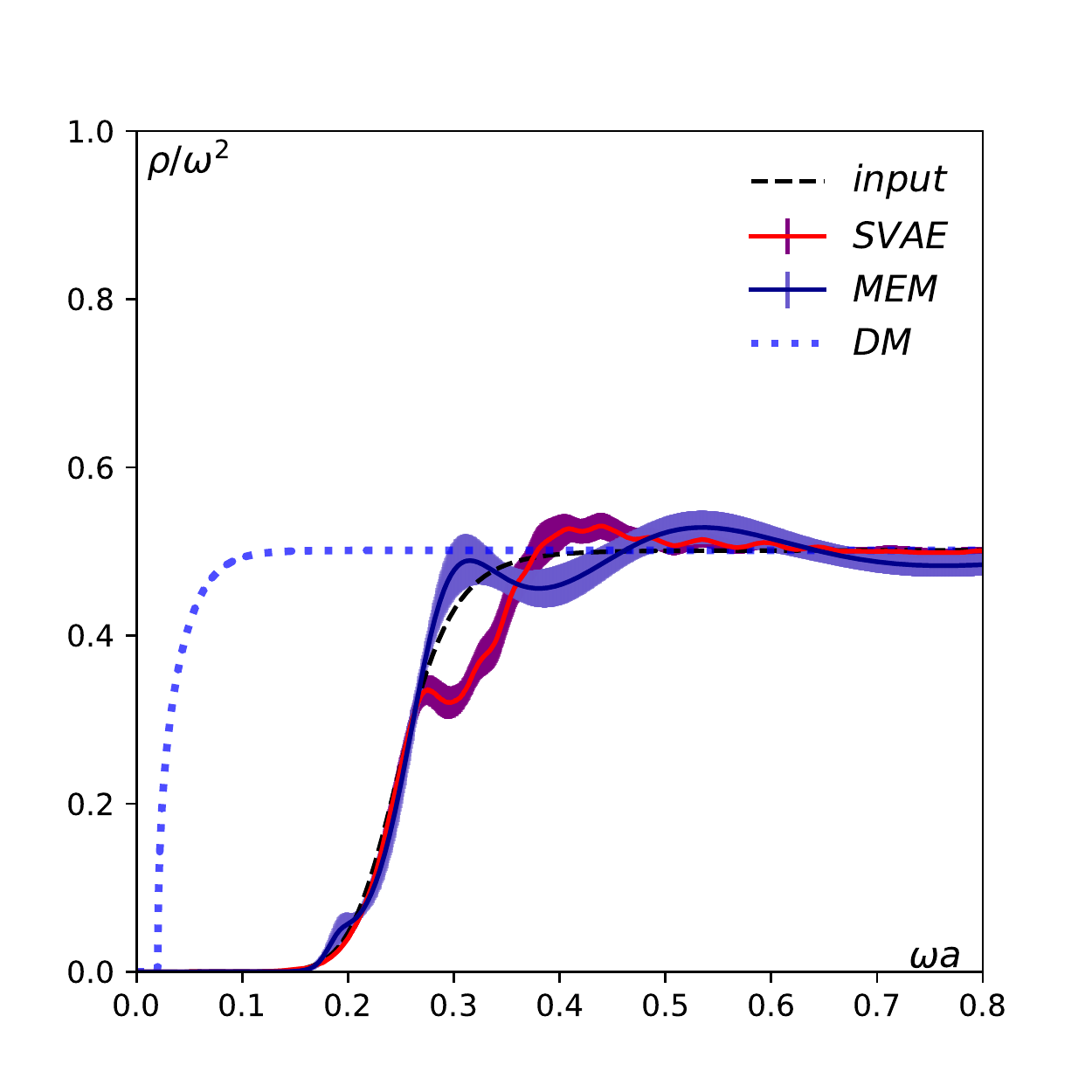}
		\includegraphics[width=0.45\textwidth]{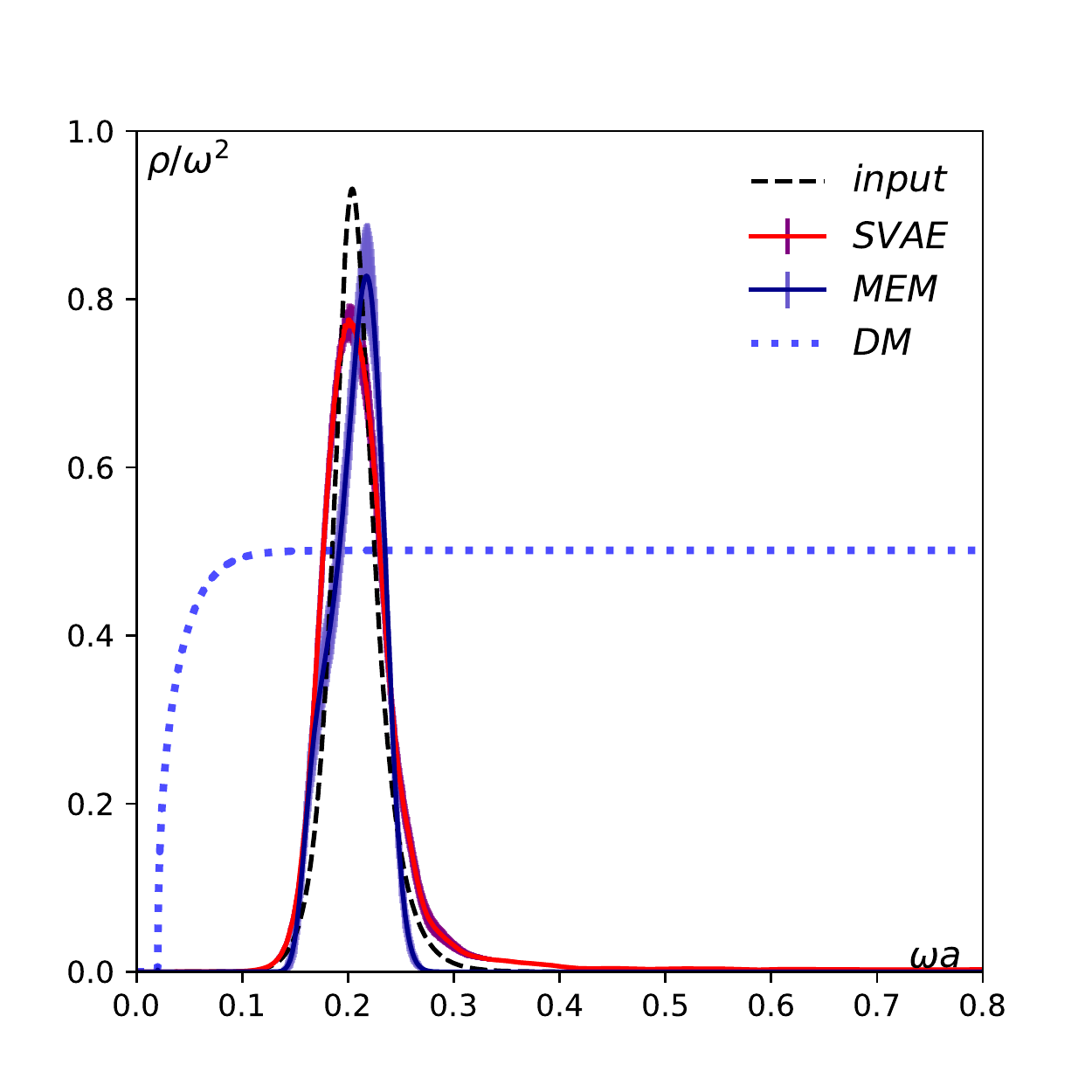}
	\caption{Left: Mock data tests with an input spectral function containing only a continuum part (cf. Eq.~\eqref{eq:res+cont} with $C_{res}=0$). Right: Same as the left one but with an input spectral function containing only a resonance peak (cf. Eq.~\eqref{eq:res+cont} with $C_{cont}=0$). Results are obtained with $N_{\tau}=96$. }
	\label{fig:case12_Nt96}
\end{figure}
We start by showing the results obtained in the simplest cases, i.e. with only a continuum part (left) and only a resonance peak (right) in the input spectral functions in Fig.~\ref{fig:case12_Nt96}. The former corresponds to the case in the non-interacting high temperature limit, while the latter corresponds to an idea case with only one ground state resonance peak in the spectral function. For the input mock spectral function shown in the left plot we use the form of Eq.~\eqref{eq:res+cont} with $C_{res}=0$ while for that shown in the right plot we use the form of Eq.~\eqref{eq:res+cont} with  $C_{cont}=0$. In the tests hereafter, we fix $M_{cont}$ to be 0.05 in Eq.~\eqref{eq:res+cont}. The results are obtained with $N_\tau=96$. The input mock spectral function is denoted by the black dashed line, and the mean value and uncertainty of the output spectral function from the trained SVAE are represented by the red solid line and purple band, respectively. For comparison hereafter we also include the output spectral function obtained from the MEM. A simple free continuum spectral function is chosen as the default model to avoid the complexity in the MEM analysis and it is denoted by blue dotted line. The mean value and uncertainty of the spectral function from MEM are denoted by the black solid line and blue band, respectively. It can be observed that both output spectral functions from the SVAE and MEM can more or less reproduce the input spectral functions. In the case of reproducing the peak location of the resonance peak the SVAE even outperforms MEM. We also check results with $N_\tau=48$, and find that the output spectral function from the SVAE and MEM are comparable with the inputs.

\begin{figure}[!htp]
	\includegraphics[width=\textwidth]{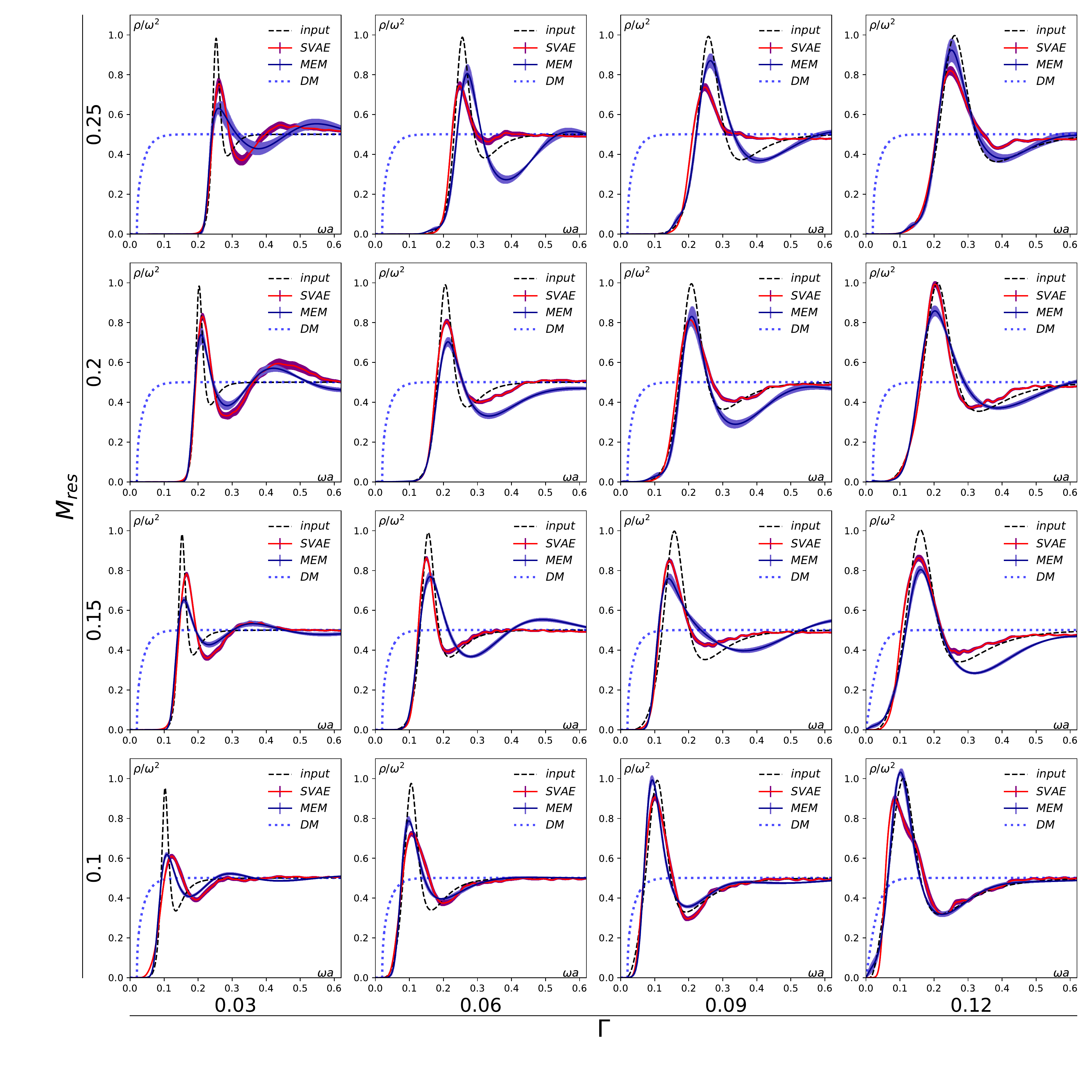}
	\caption{Mock data test with input spectral function containing a resonance peak and a continuum part (cf. Eq.~\eqref{eq:res+cont} with $N_{\tau}=96$. The black dashed line denotes the input spectral function. From left to right the width of the resonance peak in the input spectral function, $\Gamma$, is increased with the peak location of the resonance peak, $M_{res}$, fixed in the each row, while from bottom to top $M_{res}$ is increased with $\Gamma$ fixed in the each column.  The red solid line and purple band represent the mean values and uncertainties of spectral functions reconstructed from the SVAE, respectively. The black solid line and blue band denote the mean values and uncertainties of spectral functions reconstructed from the MEM with the blue dotted line the default model. }
	\label{fig:case3_Nt96}
\end{figure}

\begin{figure}[!htp]
	\centering
	\includegraphics[width=\textwidth]{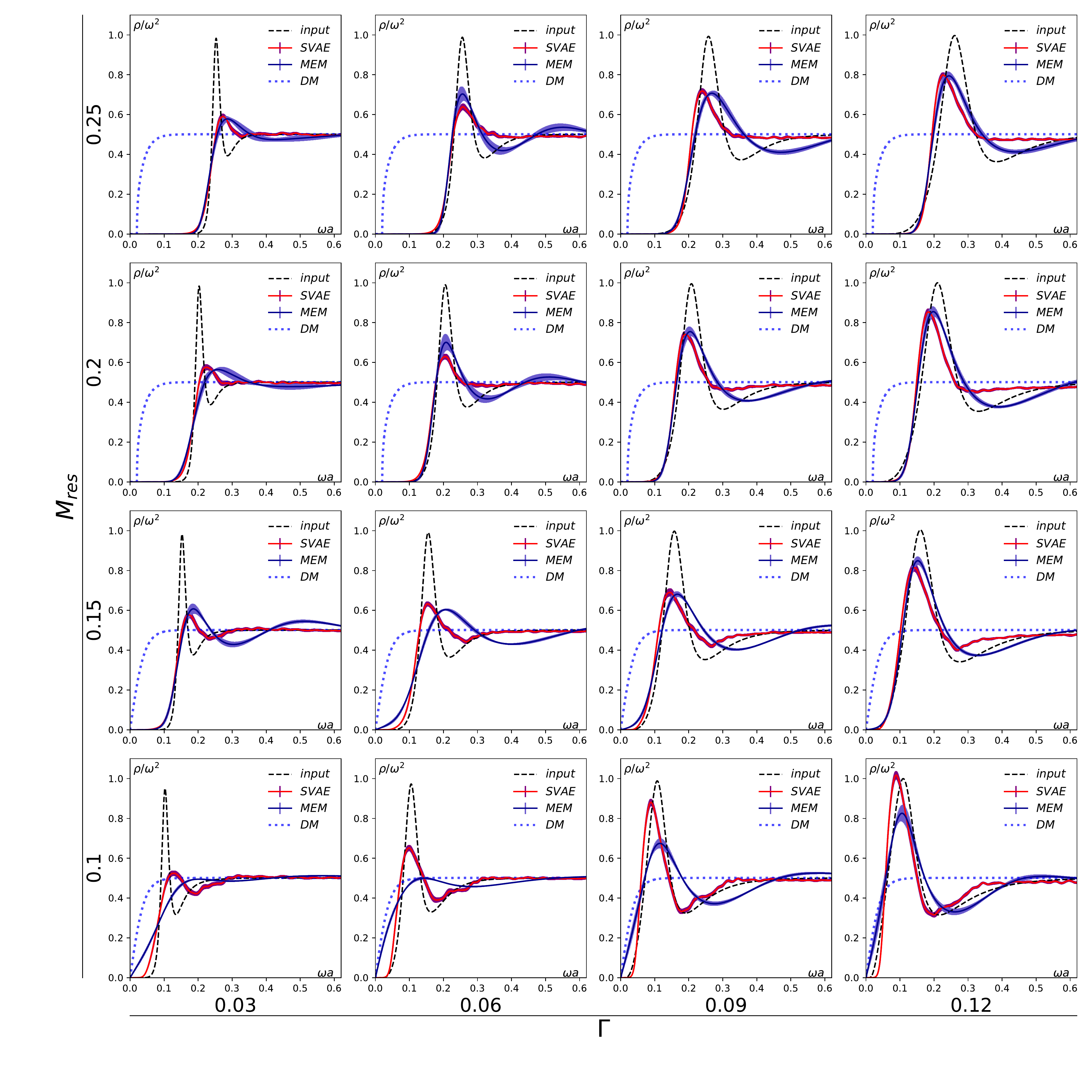}
	\caption{Same as Fig.~\ref{fig:case3_Nt96} but with $N_\tau$=48.  }
	\label{fig:case3_Nt48}
\end{figure}

Next, we show the test results with a more realistic spectral function in the interacting case as the input in Fig.~\ref{fig:case3_Nt96}. This input mock spectral function consists of a resonance peak and a continuum part, i.e. $\rho_{res+con}$ parametrized in Eq.~\eqref{eq:res+cont}. The test results shown in Fig.~\ref{fig:case3_Nt96} are done with $N_\tau=96$. In this test we focus on the dependence of the reconstructed spectral functions on the peak location ($M_{res}$) and width ($\Gamma$) of the input $\rho_{res+con}$. Thus here in the parametrization of $\rho_{res+con}$ in  Eq.~\eqref{eq:res+cont}, $C_{res}$ and $C_{cont}$ are fixed to be 2 and 2.1, respectively, and the peak locations as well as peak width are varied for tests. Four values of $M_{res}$=0.1, 0.15, 0.2 and 0.25 are chosen, in which $M_{res}=0.15$ corresponds to the mass of $\eta_c$. That is we have one value below $M_{\eta_c}$, one equal to $M_{\eta_c}$ and two above $M_{\eta_c}$ with $M_{\eta_c}$ being the mass of $\eta_c$. The smallest value of peak width $\Gamma$ is 0.03, corresponding to $\sim$60 MeV in physical units while the largest value of $\Gamma$ is four times the smallest one, corresponding to $\sim$240 MeV. In each row of Fig.~\ref{fig:case3_Nt96} the value of $M_{res}$ is fixed while the value of $\Gamma$  in the input spectral function is increased from left to right. In each column the value of $\Gamma$ is fixed while the value of $M_{res}$ is increased from bottom to top. The input spectral function is denoted by the black dashed line, while the reconstructed spectral functions obtained from the SVAE and MEM are represented by the red and black solid lines, respectively. In most cases the reconstructed spectral function from the SVAE are quite comparable to that from the MEM. In general the peak location is better reproduced than the peak height.  For instance at $\Gamma=0.03$ and $M_{res}=0.1$ the reconstructed peak height from both the SVAE and MEM is about 60\% of that of the input spectral function, while the peak location is slightly better reproduced by the MEM, i.e. MEM yields a peak location at $\omega\simeq0.11$ while SVAE give a peak location at $\omega\simeq0.12$. With $\Gamma=0.03$ fixed and $M_{res}$ increased one can see that the peak height is better reproduced by the SVAE, i.e. the reconstructed peak height from the SVAE is always about 80\% of the input spectral function, while for MEM it is $\sim60\%$ at $M_{res}=0.15$ and 0.25, and $\sim75\%$ at $M_{res}=0.2$. The peak height can only be reproduced much better when $\Gamma$ becomes much larger, i.e. $\Gamma=0.12$ for all different values of $M_{res}$. This observation imposes the challenge to reconstruct sharp peaks from both the SVAE and MEM.

To check the $N_\tau$ dependence of the reconstructed spectral functions we show similar plots as Fig.~\ref{fig:case3_Nt96} in Fig.~\ref{fig:case3_Nt48} but with $N_\tau=48$. Note that the input spectral function shown in Fig.~\ref{fig:case3_Nt48} and Fig.~\ref{fig:case3_Nt96} are the same. This test on $N_\tau=48$ is important since the hadron correlator has a trivial temperature dependence coming from the integral kernel $K(\omega,\tau)$ (cf. Eq.~\eqref{eq:Grho} and Eq.~\eqref{eq:kernel}) even given that the spectral function remains unchanged by the thermal effects.
In general the spectral function is better reconstructed from the SVAE than MEM in particular when $M_{res}$ and $\Gamma$ are small, and in both cases the quality of the reconstruction is worse compared to those obtained with $N_\tau=96$. At $M_{res}=0.1$ and $\Gamma=0.03$ the reconstructed spectral function from the SVAE has a peak location that is about 30\% larger and a peak height close to $60\%$ as compared to that of the input spectral function. On the other hand, the reconstructed spectral function from MEM does not even possess a peak structure. As one increases $\Gamma$, a marked peak structure obtained from the MEM starts to be seen only at $\Gamma=0.09$ and the reconstructed resonance peak becomes comparable with that from the SVAE. On the other hand, when $\Gamma$ is fixed the reconstruction quality of the spectral function seems from the SVAE to be insensitive on $M_{res}$. This is similar as the case with $N_{\tau}=96$ as shown in Fig.~\ref{fig:case3_Nt96}.

To check the dependences of output spectral functions shown in Figs.~\ref{fig:case3_Nt96} and~\ref{fig:case3_Nt48} on the hyperparameters of the SVAE, we performed investigations on random seeds in generating training samples, number of resonances ($\hat{N_g}$) used in the training samples, number of hidden layers in the encoder and decoder and number of training epoches. We find that the dependences are mild. I.e. the features we observed in Figs.~\ref{fig:case3_Nt96} and~\ref{fig:case3_Nt48} are stable and reliable. In the following we thus adopt the hyperparameters and training parameters listed in Tables~\ref{tab:hyperpara} and ~\ref{tab:traning_window} through our study. Details of these investigations can be found in Appendix~\ref{app:dep}.

 In the previous results shown in our study we used the same Gaussian noise in the correlator for training and tests. We then also investigated the dependences on the noise model used in the mock correlator data. We used a log-normal noise~\cite{DeGrand:2012ik} in the correlator for both training and test procceses. Together with the Gaussian noise, we thus have four different combinations: either Gaussian noise or log-normal is used in the training or test process. We found that results of spectral functions obtained in following cases have minor difference compared to our current results with Gaussian noise used in both training and tests. These cases are: 1) the log-normal noise is used in both the training and tests; 2) a mismatch in the noise model: Gaussian noise is used in the training while log-normal noise is used in the tests and vice versa and 3) the multivariate Gaussian noise, i.e. including correlations among different time slices, used for the test correlator data, while the Gaussian noise is used for the training. The details on these investigation are shown in Appendix~\ref{app:noise_model}. Thus in the following studies we will only adopt the Gaussian noise model.

\begin{figure}[!htp]
	\centering
	\includegraphics[width=0.4\textwidth]{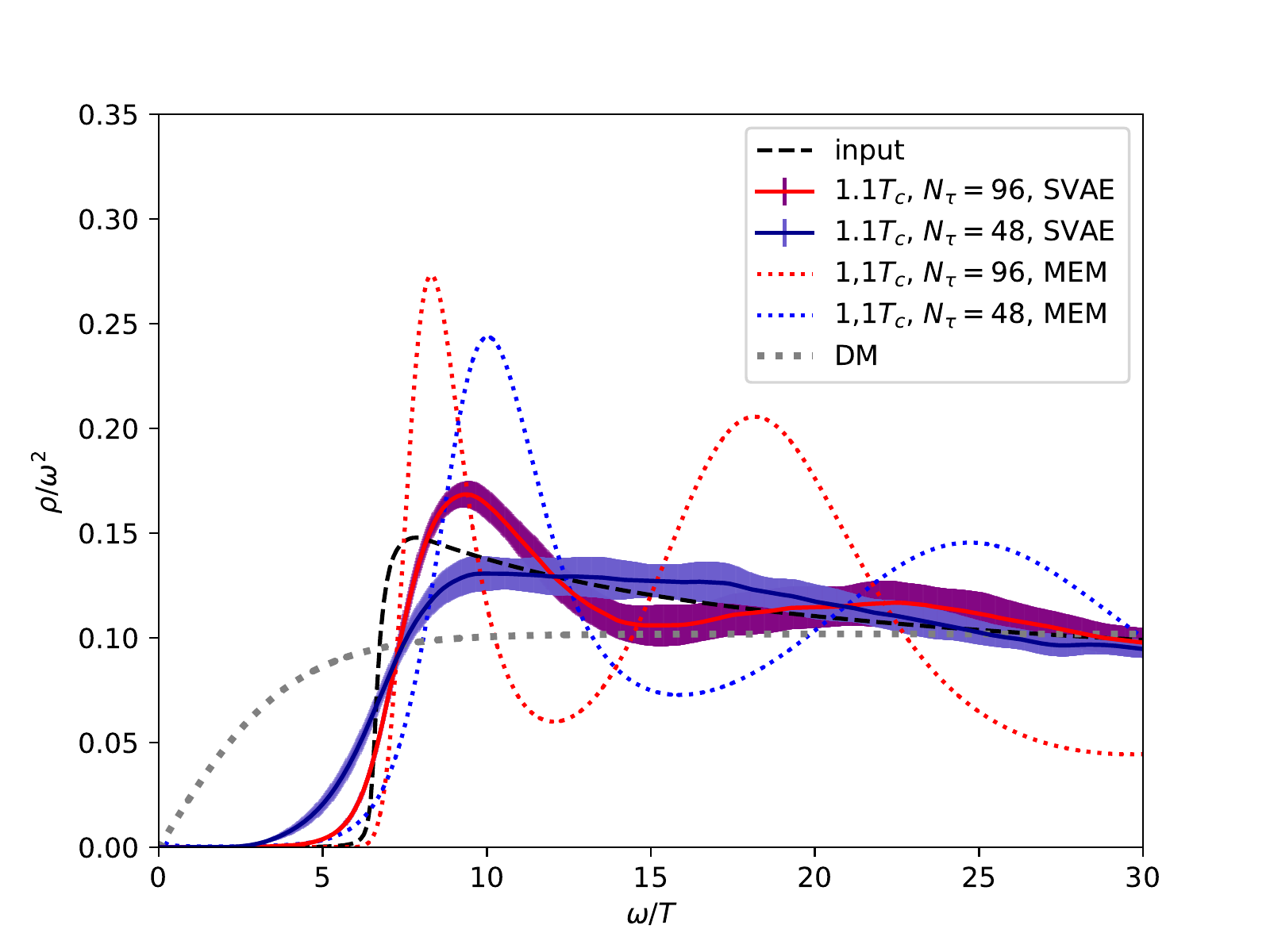}
	\includegraphics[width=0.4\textwidth]{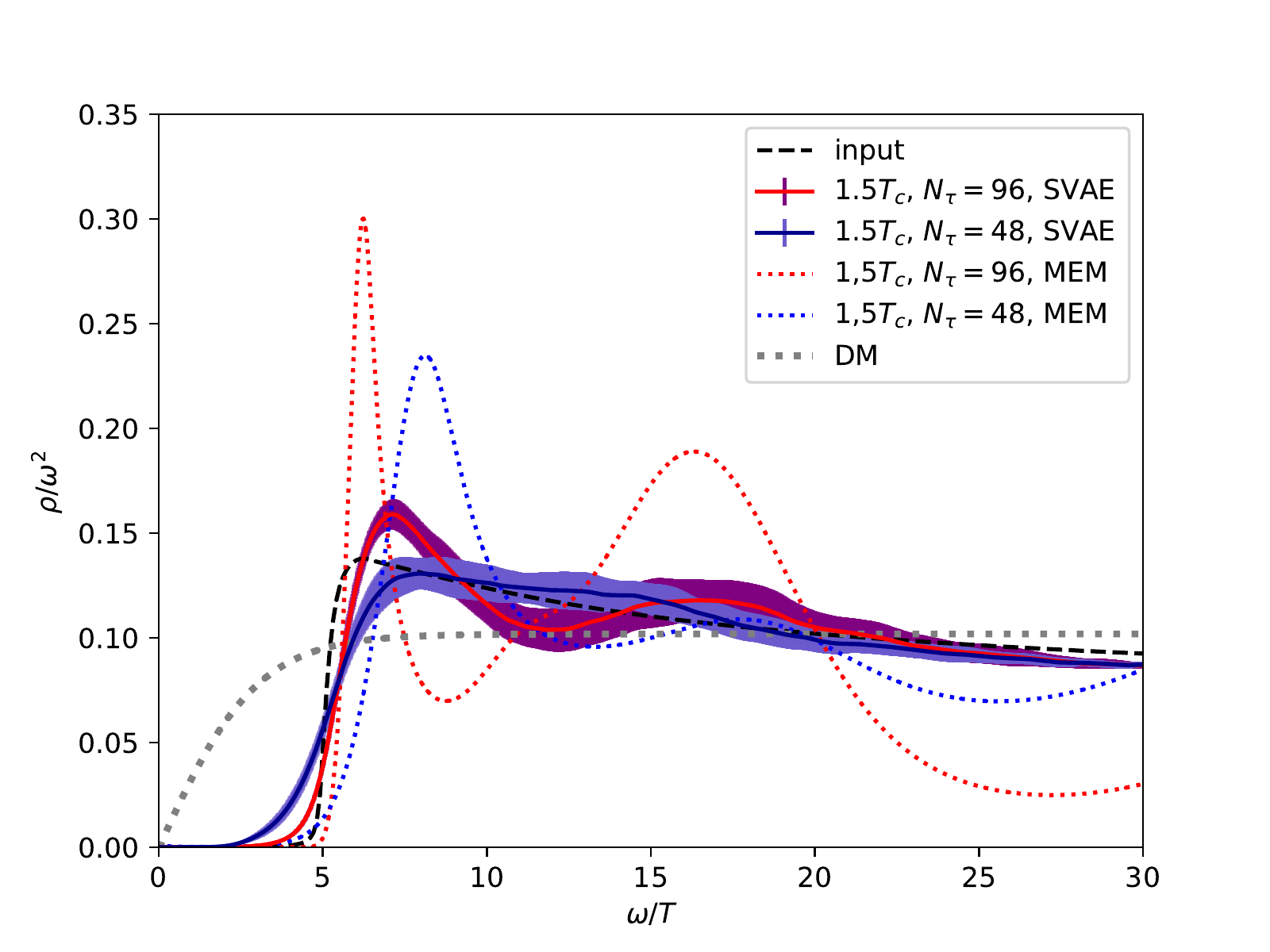}
	\caption{Mock data tests with input spectral functions obtained from 2-loop perturbative NRQCD (cf. Eq.~\eqref{eq:pertspf}) at T=1.1$T_c$ (left) and 1.5$T_c$ (right) with $N_\tau = 96$ and 48. In both plots the input spectral function is denoted by the black dashed line, and the red solid line and blue solid line with surrounding bands denote the results obtained from the SVAE with $N_\tau=96$ and 48, respectively. The dotted lines with the same color denote results obtained from the MEM with the default model denoted as grey dotted lines. 
	}
	\label{fig:case4NRQCD}
\end{figure}

We make further tests using the input spectral function obtained using pNRQCD at 1.1 $T_c$ (left) and 1.5 $T_c$ (right) with $N_\tau=48$ and 96 in Fig.~\ref{fig:case4NRQCD}. In this case there are no marked peak structures in the input spectral functions at both 1.1 and 1.5 $T_c$, and both input spectral functions arise rapidly in the low frequency part and then slowly decrease in $\omega$ after reaching its maximum value. As seen from the left plot of Fig.~\ref{fig:case4NRQCD}, the reconstructed spectral function from the SVAE with $N_\tau=48$ seems to be able to resemble the feature that the input spectral function does not possess a peak structure, and shifts the location of the maximum of the spectral function to a larger value of $\omega/T$. On the other hand, the MEM gives a wrong result even at the qualitative level, i.e. it shows a fake pronounced peak structure at $N_\tau=48$. When going to a larger value of $N_\tau$, i.e. 96, in the case of SVAE the location of the maximum remains almost unchanged and the maximum surpasses that of the input spectral function, while in the case of MEM, the location of the maximum becomes closer to that of the input and the peak becomes sharper. Similar features of reconstructed spectral functions can be seen in the right plot of Fig.~\ref{fig:case4NRQCD} having a pNRQCD spectral function at 1.5 $T_c$ as the input. Thus it seems that both the SVAE and MEM cannot reproduce the input pNRQCD spectral function in a satisfactory way.

We close this section by comparing the SVAE and MEM. In most cases spectral functions obtained from the SVAE are comparable to those obtained from the MEM, e.g. in the mock data tests shown in Figs.~\ref{fig:case12_Nt96},~\ref{fig:case3_Nt96} and even outperform the MEM in reconstructing sharp peaks with $N_\tau=48$, see e.g. Fig.~\ref{fig:case3_Nt48}. In the case with pNQRCD spectral funciton as inputs both the SVAE and MEM cannot reproduce the input spectral function well. In this case a sharp resonance peak is obtained from the MEM and a bump structure is found from the SVAE, although there is no marked peak structure in the inputs as seen from Fig.~\ref{fig:case4NRQCD}. These tests thus implies the importance of studies of spectral function using different methods. In the next session we will apply the SVAE to lattice QCD data, and also use the MEM as a complementary method to study the systematic uncertainties.

\section{Analyses on lattice QCD data}
\label{sec:lqcd}

In this section we will present the charmonia spectral functions in the pseudo-scalar ($\eta_c$) channel extracted using the proposed SVAE and MEM. The lattice QCD correlators used in the current analyses are taken from Ref.~\cite{Ding:2012sp} and here we focus on the correlators computed on the finest lattices, i.e.  $128^3\times96$ and $128^3\times48$ corresponding to temperature at $0.75~T_c$ and $1.5~T_c$ with inverse lattice spacing $a^{-1}=18.97$ GeV. The mass of $\eta_c$ extracted from the exponential decay of temporal correlation function in this lattice setup is 3.341(2)(104) GeV  with the error in the first bracket statistical error and the error in the second bracket systematic error from effects of the physical distance~\cite{Ding:2012sp}. As the correlators calculated on lattices suffer from the lattice cut-off effects, which would manifest themselves mostly at small distances or large $\omega$, we thus abandon the first four points of the correlators in the short distance in our analyses as was done in the mock data tests. In the following MEM analyses we adopt three default models: ``DM1" contains only free spectral functions, ``DM2" consists a continuum part and a resonance peak with peak location smaller than the $\eta_c$ mass in the vacuum, and ``DM3" is similar as ``DM2" but with the peak location larger than the $\eta_c$ mass in the vacuum.

\begin{figure}[!htp]
	\centering
	\includegraphics[width=0.4\textwidth]{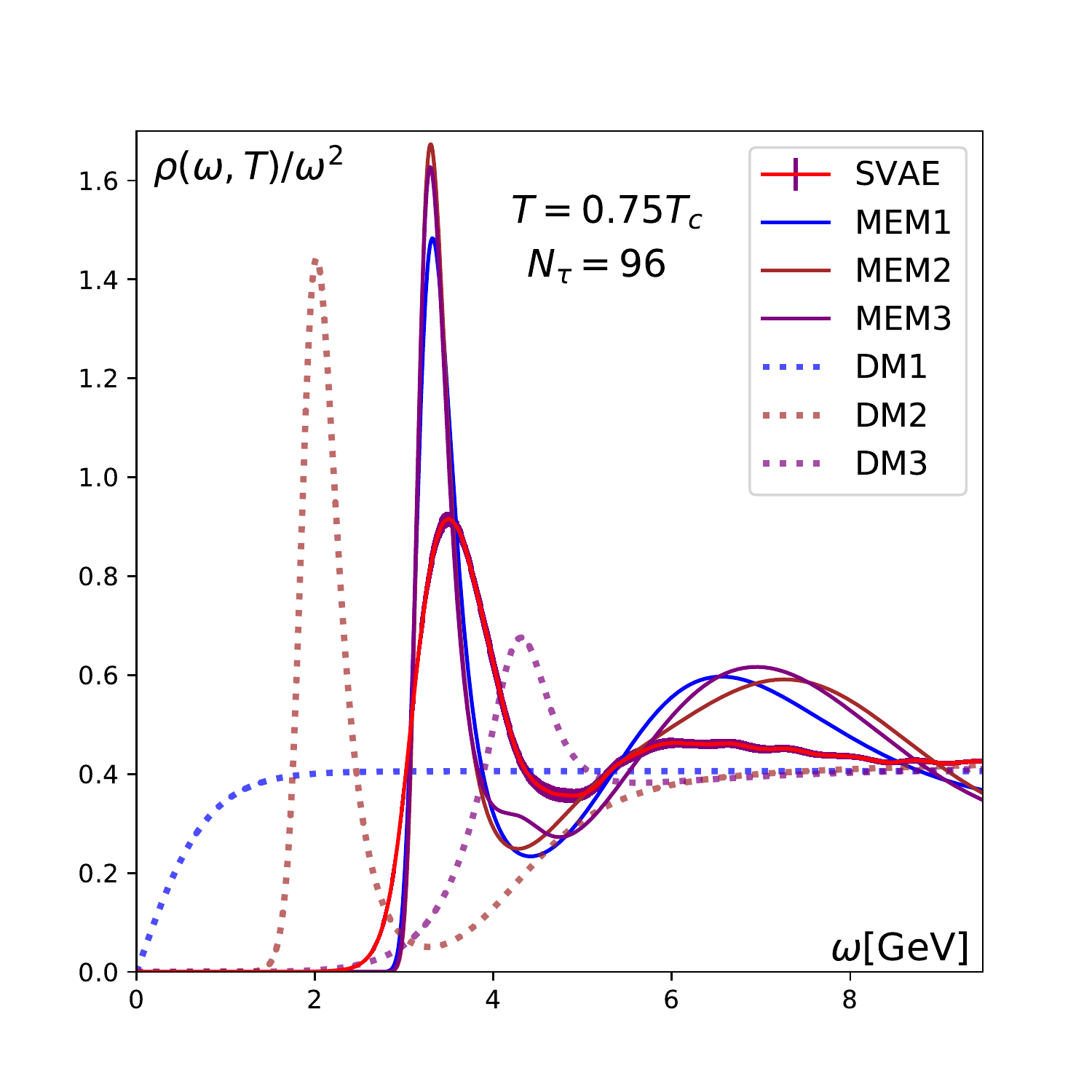}
	\includegraphics[width=0.4\textwidth]{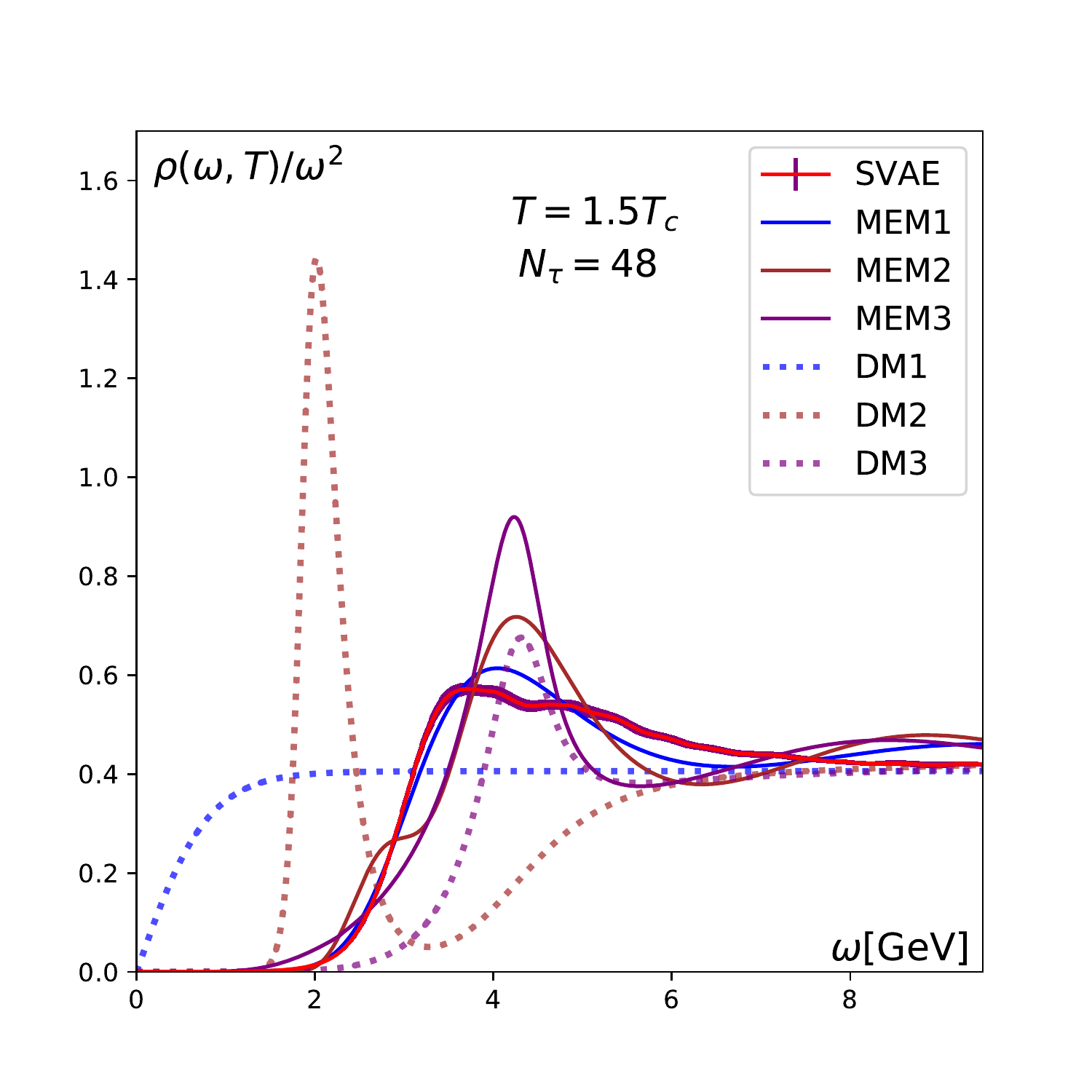}
	\caption{Spectral functions extracted from temporal correlation functions computed in LQCD simulations~\cite{Ding:2012sp} with SVAE and MEM at 0.75 $T_c$ (left) and 1.5 $T_c$ (right).  In both plots, the output spectral functions from the SVAE are denoted by the red solid line and their uncertainties are represented by the surrounding purple band, while the outputs from the MEM are denoted by the solid lines while the dotted line with the same color are the corresponding default models used in the MEM analyses.}. 
\label{fig:lattice_spf}
\end{figure}

In left plot of Fig.~\ref{fig:lattice_spf} we show the reconstructed spectral function (solid lines) using the SVAE and MEM from lattice data at 0.75 $T_c$ with $N_\tau=96$. The peak location and peak height of the reconstructed spectral function from the SVAE are at $\simeq$ 3.47 GeV and about 0.9, respectively~\footnote{Hereafter the peak we refer to is always the first peak starting from the low $\omega$. The second peak normally comes from the lattice artefacts~\cite{Ding:2012sp}.}. The obtained peak location corresponding to the mass of $\eta_c$ is consistent with that extracted from the exponential decay of the temporal correlation functions, i.e. 3.341(2)(104) GeV~\cite{Ding:2012sp}. On the other hand, the spectral function obtained from MEM show small default model dependences, and the reconstructed peak location lies at $\simeq3.34$ GeV with the peak height about 1.5-1.7. Thus the peak location reconstructed from SVAE and MEM are comparable with each other and differ by less than 5\%, and the peak height obtained from MEM is about 66-88\% larger than that from the SVAE. This suggests that the peak locations reconstructed from these two methods are more reliable compared to the peak height. This is consistent with the mock data tests, i.e. the peak locations obtained using these two methods are comparable as seen from e.g. Fig.~\ref{fig:case3_Nt96} and the peak height from MEM is much larger than that from the SVAE as seen from Fig.~\ref{fig:case4NRQCD}.

In the right plot of Fig.~\ref{fig:lattice_spf} we show the reconstructed spectral function obtained from lattice data at 1.5 $T_c$ with $N_\tau=48$. The peak location reconstructed from the SVAE shifts to $\omega\simeq3.72$ GeV and that from the MEM shifts to $\omega\simeq4.05$ GeV with ``DM1" and $\omega\simeq4.17$ GeV using ``DM2" and ``DM3".  The peak height obtained from the SVAE, on the other hand, shrinks to about $65\%$ of its value at 0.75 $T_c$. For the MEM results the peak height becomes $\simeq$40\% and $\simeq54\%$ of their corresponding values at 0.75$T_c$ by using ``DM1" or ``DM3", and ``DM2", respectively. The general structure of output spectral function from the SVAE is similar to that from the MEM using the default model without a resonance peak, i.e. ``DM1". Based on the results obtained from both methods it seems that the resonance peak for $\eta_c$ is substantially modified at 1.5 $T_c$.

Since at higher temperatures the information of the spectral function is compressed into the correlation function with a short temporal extent $aN_\tau$ due to $T=1/(aN_\tau)$, we further study the uncertainties arising from the smaller temporal extent at 1.5$T_c$. To do so we look at the ``reconstructed" correlator at temperature $T$ from a spectral function determined at temperature $T^{\prime}$
\begin{align}
G_{rec}(\tau,T;T^{\prime})= \int_{0}^{\infty} \mathrm{d}\omega ~K(\tau,T,\omega)~\rho(\omega,T^{\prime}).
\end{align}
$G_{rec}(\tau,T;T^{\prime})$ at  temperature $T$ can be computed from a spectral function $\rho(\omega,T^{\prime})$ at  temperature $T^{\prime}$ and an integral kernel $K$ at  temperature $T$. One actually can avoid computing $\rho(\omega,T^{\prime})$ to arrive at $G_{rec}(\tau,T;T^{\prime})$, and can compute $G_{rec}(\tau,T;T^{\prime})$ directly based on the correlator at a temperature $T^{\prime}$ as follows~\cite{Ding:2012sp}
\begin{align}
G_{rec}({\tau},T;T^{\prime}) = \sum_{{\tau}^{\prime}={\tau};~\Delta{\tau}^{\prime}=N_{\tau}}^{N_{\tau}^{\prime}-N_{\tau}+{\tau}}  G({\tau}^{\prime},T^{\prime}) \,,
\label{eq:Grec_data}
\end{align}
where $T^{\prime}=(a N_{\tau}^{\prime})^{-1},~~T=(aN_{\tau})^{-1},~~\tilde{\tau}^{\prime}=(\tau^{\prime}/a)\in[0,~N_{\tau}^{\prime}-1],~~\tilde{\tau}=(\tau/a)\in[0,~N_{\tau}-1],~~N_{\tau}^{\prime}=m~ N_{\tau},~~m\in\mathbb{Z}^{+}$. $N_{\tau}$ and $N_{\tau}^{\prime}$ are the number of time slices in the temporal direction at temperature $T$ and $T^{\prime}$, respectively; $\tilde{\tau}$ denotes the time slice of the correlation function at temperature $T$ while $\tilde{\tau}^{\prime}$ represents the time slice of the correlation function at 
temperature $T^{\prime}$. The sum over $\tilde{\tau}^{\prime}$ on the right hand side of ~\eqref{eq:Grec_data} starts from $\tilde{\tau}^{\prime}=\tilde{\tau}$ with a step length of $\Delta\tilde{\tau}^{\prime}=N_{\tau}$ and ends at the upper limit $N_{\tau}^{\prime}-N_{\tau}+\tilde{\tau}$.
Note that the spectral function encoded in the correlators on both sides of the above equation is the same. Using relation~\eqref{eq:Grec_data} we can calculate $G_{rec}(\tau ,1.5T_c;0.75T_c)$ having $N_\tau=48$ directly from the correlator data at temperature $T^{\prime}=0.75T_c$ with $N_\tau=96$ and a step size $\Delta{\tau}^{\prime}$. We thus can study the $N_\tau$ dependence of output spectral function $\rho(\omega,0.75T_c)$ obtained with the SVAE and MEM. 

\begin{figure}[!htp]
	\centering
		\includegraphics[width=0.4\textwidth]{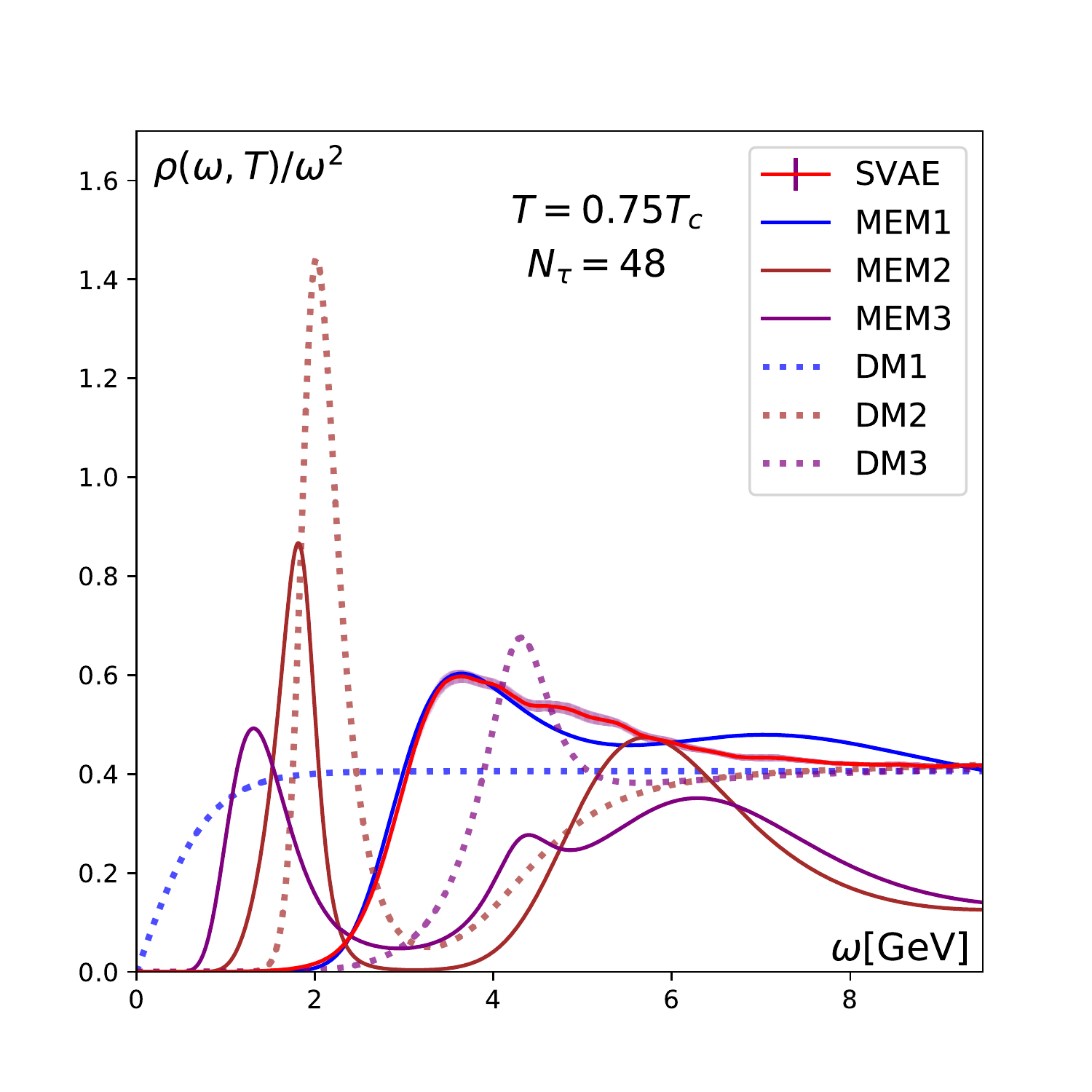}
	\includegraphics[width=0.4\textwidth]{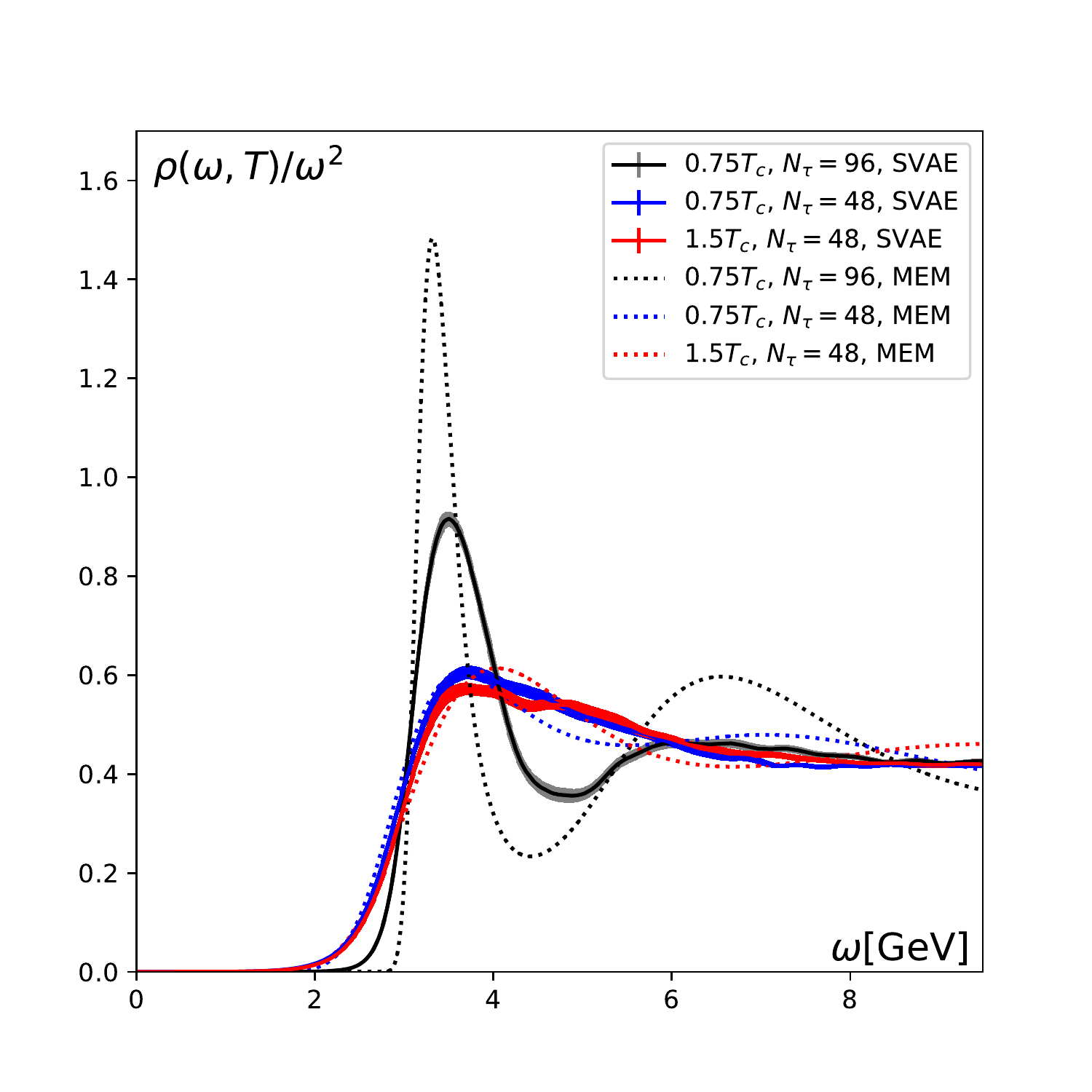}
	\caption{Left: Spectral functions extracted from the reconstructed correlator where $N_\tau=48$ and the encoded spectral function should be that at 0.75 $T_c$. Right: Spectral function at 0.75 $T_c$ and 1.5 $T_c$ obtained from the SVAE and VAE based on $N_\tau=48$ and $96$ correlator data. Only results obtained using ``DM1" in MEM analyses are shown.} 
	\label{fig:lqcd_uncertainty}
\end{figure}
In the left plot of Fig.~\ref{fig:lqcd_uncertainty} we show the output spectral functions obtained using the SVAE and MEM based on the reconstructed lattice data. The spectral function encoded in the reconstructed correlator data should be that at 0.75 $T_c$, however, the correlator data has $N_\tau=48$. For the spectral function obtained from the SVAE it can be clearly seen that the peak obtained with $N_\tau=48$ becomes much broader compared to the one obtained with $N_\tau=96$ (cf. the red solid line shown in the left plot of Fig.~\ref{fig:lattice_spf}), and the peak location shifts from $\omega\simeq$ 3.47 GeV to $\omega\simeq$ 3.72 GeV while the peak height reduces from about 0.9 to 0.6. On the other hand, a large default model dependence is found in the output spectral function obtained from the MEM. It can also be observed that the output spectral function from MEM using ``DM1" is comparable to that obtained using the SVAE. Thus the large $N_\tau$ dependence is found in both SVAE and MEM during the extraction of $\rho(\omega,0.75T_c)$. This suggests that $N_\tau$ larger than 48 is needed to resolve the fate of $\eta_c$ at 1.5 $T_c$.

In the right plot of Fig.~\ref{fig:lqcd_uncertainty} we show a summary of the output spectral function obtained using the SVAE and MEM. The results obtained from MEM are shown only for those obtained using a default model having no resonance peak, i.e. ``DM1". We can clearly see that the $\rho(\omega,0.75T_c)$ extracted from the reconstructed correlator data having $N_\tau=48$ is very similar to $\rho(\omega,1.5T_c)$ extracted from $N_\tau=48$ correlator at 1.5 $T_c$. Thus from the current data it is hardly able to tell whether $\eta_c$ is dissociated or not at 1.5 $T_c$. 

\section{Conclusion}
\label{sec:con}
In this study we have explored the reconstruction of spectral functions from Euclidean two-point correlation functions using machine learning. To do so we proposed a novel neural network approach based on the variational autoencoder, SVAE, where `S' stands for the entropy term $S$ (Eq.~\eqref{eq:entropy}) included in the loss function. The variational autoencoder consists of an encoder and a decoder which are connected via the latent variable $z$. Based on the Bayesian theorem, we transform the problem of specifying the posterior probability distribution $P(\rho|G)$ in the loss function to the problem of specifying the likelihood function $P(G|\rho,z)$ and the prior information $P(\rho|z)$ (represented by a Shannon-Jaynes entropy term) as well as the KL divergence of the posterior probability distribution $KL\big(Q(z|\rho_{gt},G_{gt})\| P(z|G) \big)$. Using the constructed loss function we trained the SVAE by feeding the spectral functions having general forms obtained from the Gaussian mixture model. With the trained SVAE we have tested four types of physics motivated spectral functions, i.e. spectral functions consisting of 1) one resonance peak only, 2) one continuum spectral function only, 3) one resonance peak and one continuum spectral function, and 4) spectral functions computed from pNRQCD. In these training and mock data tests we generally mimic the noise level of the lattice QCD data. In the first two cases the spectral function can be fairly reconstructed with $N_\tau=96$ and 48. For case 3) we found that the peak location of the resonance peak is better reproduced than the peak height, and as the peak width becomes narrower the harder the resonance peak can be reconstructed. A large $N_\tau$ dependence is also observed in the reconstructed spectral function. We also tested and trained the neural networks with less noisy correlators having $N_\tau=48$ and found that reducing the noise level is not as beneficial as increasing $N_\tau$ of the correlators as shown in Appendix~\ref{app:dep}.
We also checked the dependences on the noise model, i.e. Gaussian and log-normal noise, used for the mock correlator data. We found that the mismatch in the noise models gives minor differences in the output spectral functions as shown in Appendix~\ref{app:noise_model}. In these three cases the quality of the spectral reconstruction from SVAE is comparable to that from the MEM. In  case 4) both SVAE and MEM cannot reproduce the spectral function in a satisfactory way.

We have applied  SVAE to the Euclidean two-point correlators of charmonium in the pseudoscalar channel computed from quenched lattice QCD simulations on $128^3\times96$ and $128^3\times48$ lattices corresponding to 0.75$T_c$ and $1.5T_c$, respectively. With $N_\tau=96$ we find that both SVAE and MEM can reproduce the peak location of ground state peak of $\eta_c$ at 0.75$T_c$ fairly well. However, by applying the SVAE and MEM to the reconstructed correlator having $N_\tau=48$ and encoding spectral function at 0.75$T_c$ it turns out that the resonance peak is substantially modified and is comparable to the extracted spectral function at 1.5$T_c$ on $N_\tau=48$ lattices. Due to the large $N_\tau$ dependence observed in the reconstruction of spectral functions at 0.75$T_c$ it thus hardly possible to tell whether $\eta_c$ survives or not at $1.5T_c$.

As observed from the mock data tests, the unsatisfactory reconstruction of the spectral function at 0.75 $T_c$ from the reconstructed lattice correlator data with $N_\tau=48$ via both the SVAE and MEM is not a surprise. This could be due to the intrinsic ill-posed nature of the inverse problem and a large value of $N_\tau$ is needed to resolve the fate of $\eta_c$ at 1.5 $T_c$. Since the severity of the problem could change with different integral kernel in the inversion, it would be interesting in the future to apply the SVAE to correlators at zero temperature or correlators computing from NRQCD on the lattice where in both cases the integral kernel, i.e. $e^{-\omega\tau}$, does not have a trivial temperature dependence and is simpler compared to the kernel used in our current study.
Meanwhile, it could be feasible that the quality of spectral reconstruction from the SVAE can be better improved using other techniques. e.g. to have smoothness of spectral functions as required in the loss function~\cite{Wang:2021jou}.

\section*{Acknowledgement}
We thank De-Ji Liu and Akio Tomiya for the early involvement in this work, and Hai-Tao Shu on the correspondence on the MEM analyses. This work was supported by the National Key Research and Development Program of China under Contract No. 2022YFA1604900, the NSFC under the grant number 11775096, the Hungarian Research Fund (OTKA) under the grant number K123815 and the Ministry of Innovation and Technology NRDI Office within the framework of the MILAB Artificial Intelligence National Laboratory Program. The numerical simulations have been performed on the GPU cluster in the Nuclear Science Computing Center at Central China Normal University (NSC$^3$).

\begin{appendix}

\section{Dependences on the noise level of the correlator data, hyperparameters and training parameters in the SVAE}
\label{app:dep}
\begin{figure}[!hpt]
	\centering
	\includegraphics[width=0.8\textwidth]{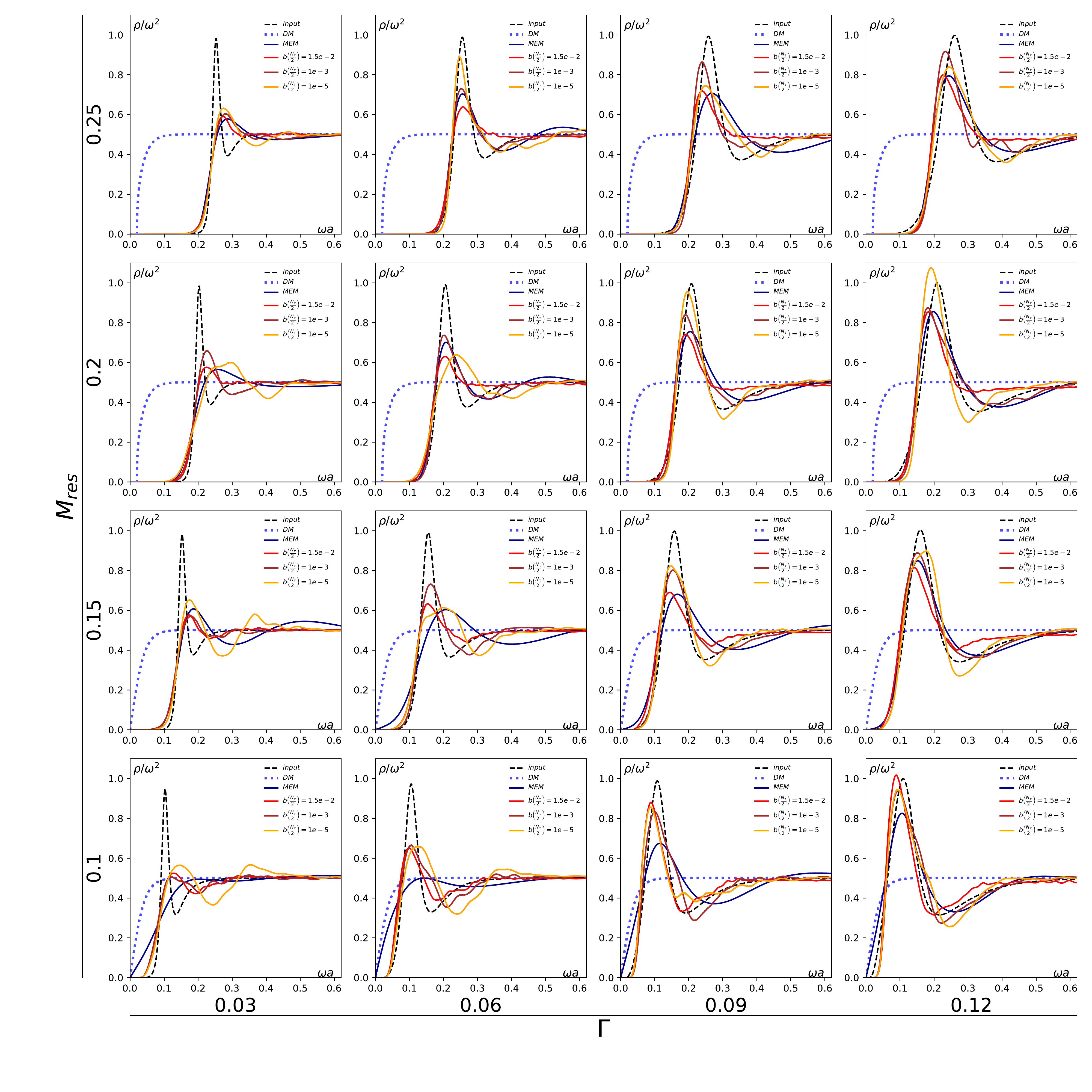}
	\caption{Spectral functions extracted from the mock correlator data with $N_\tau=48$ with different noise levels $b(\frac{N_\tau}{2})$. The input spectral functions are the same as those shown in Fig.~\ref{fig:case3_Nt96} and Fig.~\ref{fig:case3_Nt48}.}
	\label{fig:spftotal48noisedep}
\end{figure}

In this appendix we show the dependence of the output spectral function obtained using the SVAE on the noise level $b$ (cf. Eq.\eqref{eq:noise_level}) of the input correlator data. Similar study was performed for MEM in e.g.~\cite{Asakawa:2000tr}.
The noise level of the input correlator data, i.e., error dived by the mean value at $\tau=N_\tau/2$, shown in Fig.~\ref{fig:spftotal48noisedep} varies from 1.5$\times10^{-2}$ to $10^{-5}$. Note that the values of the noise level $b(\tau)$ in the training data are the same as those used in the mock data tests for consistency. It can be seen that with much precise data the benefits are not obvious. From our current test it suggests that increasing $N_\tau$ is more beneficial. 
	\begin{figure}[!hpt]
	\centering
\includegraphics[width=0.4\textwidth]{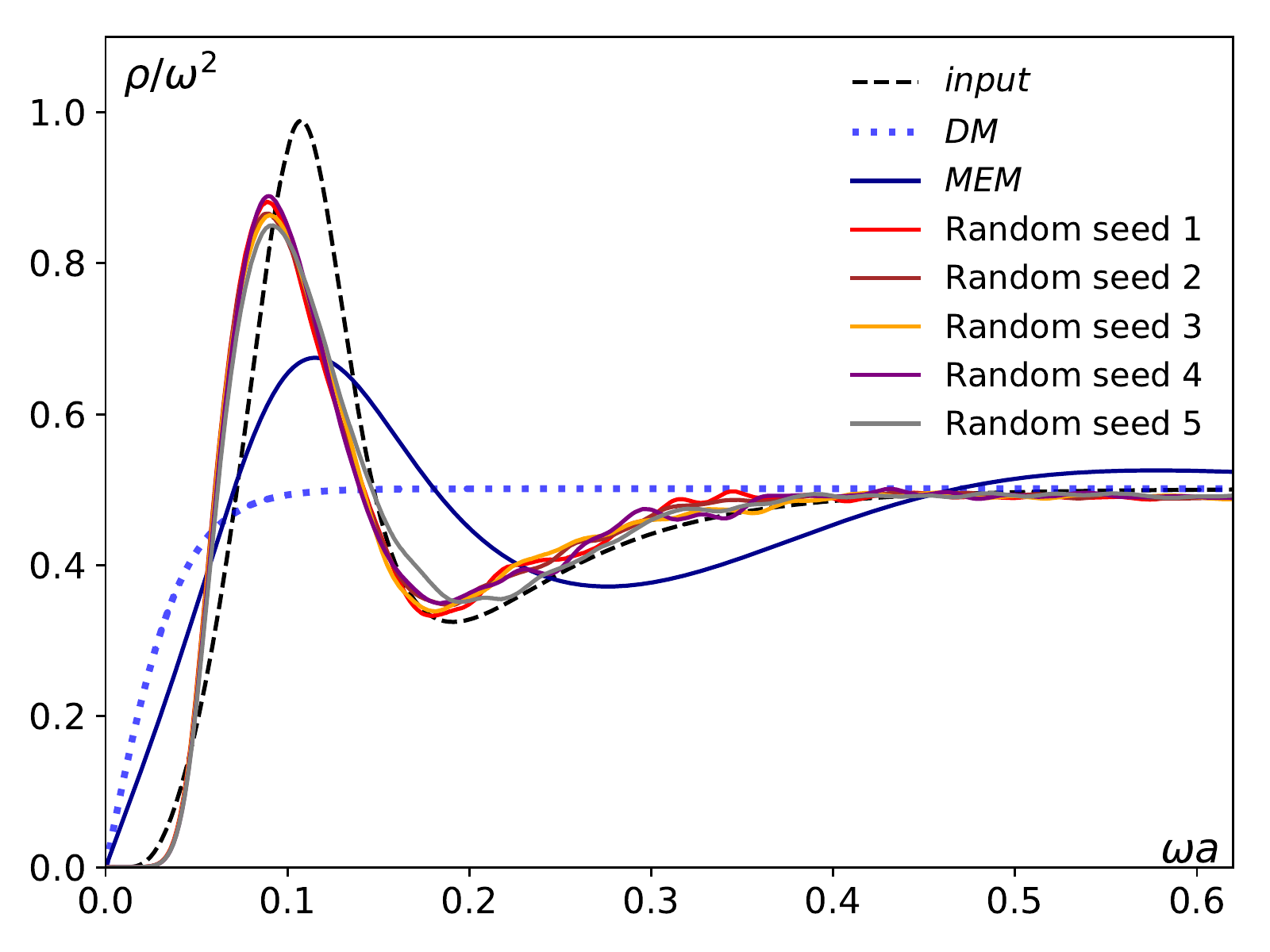}
\includegraphics[width=0.4\textwidth]{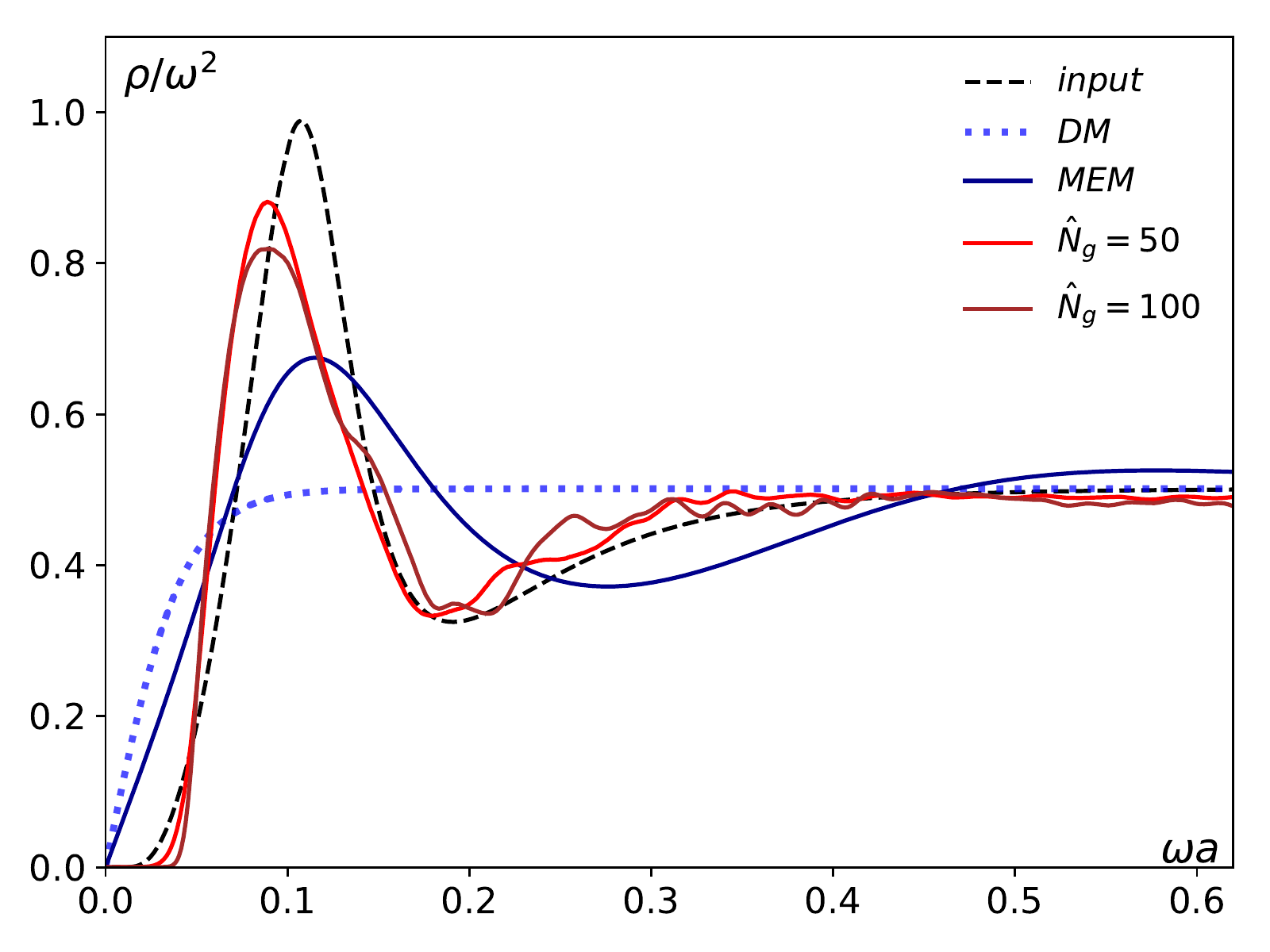}
\includegraphics[width=0.4\textwidth]{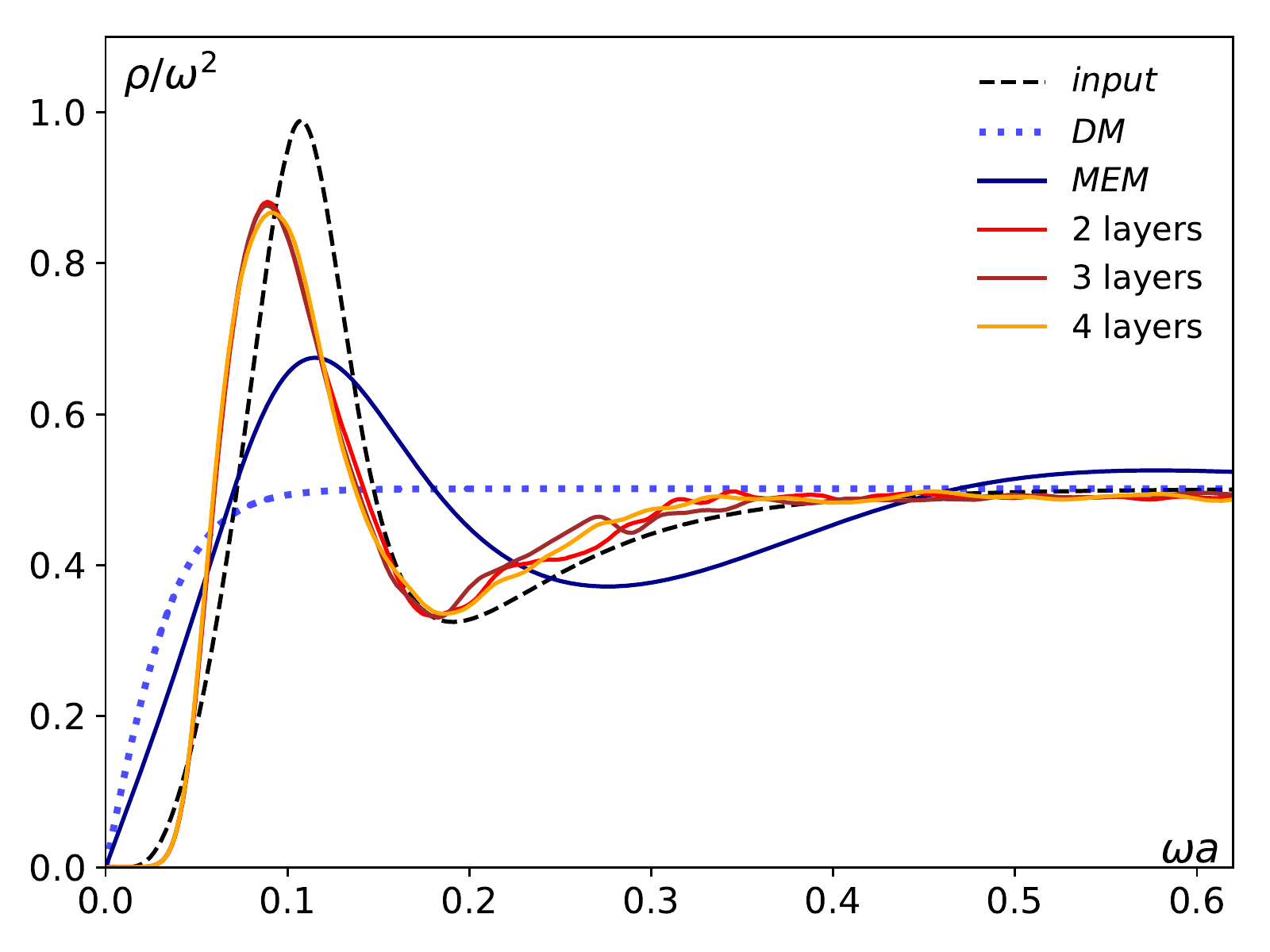}
\includegraphics[width=0.4\textwidth]{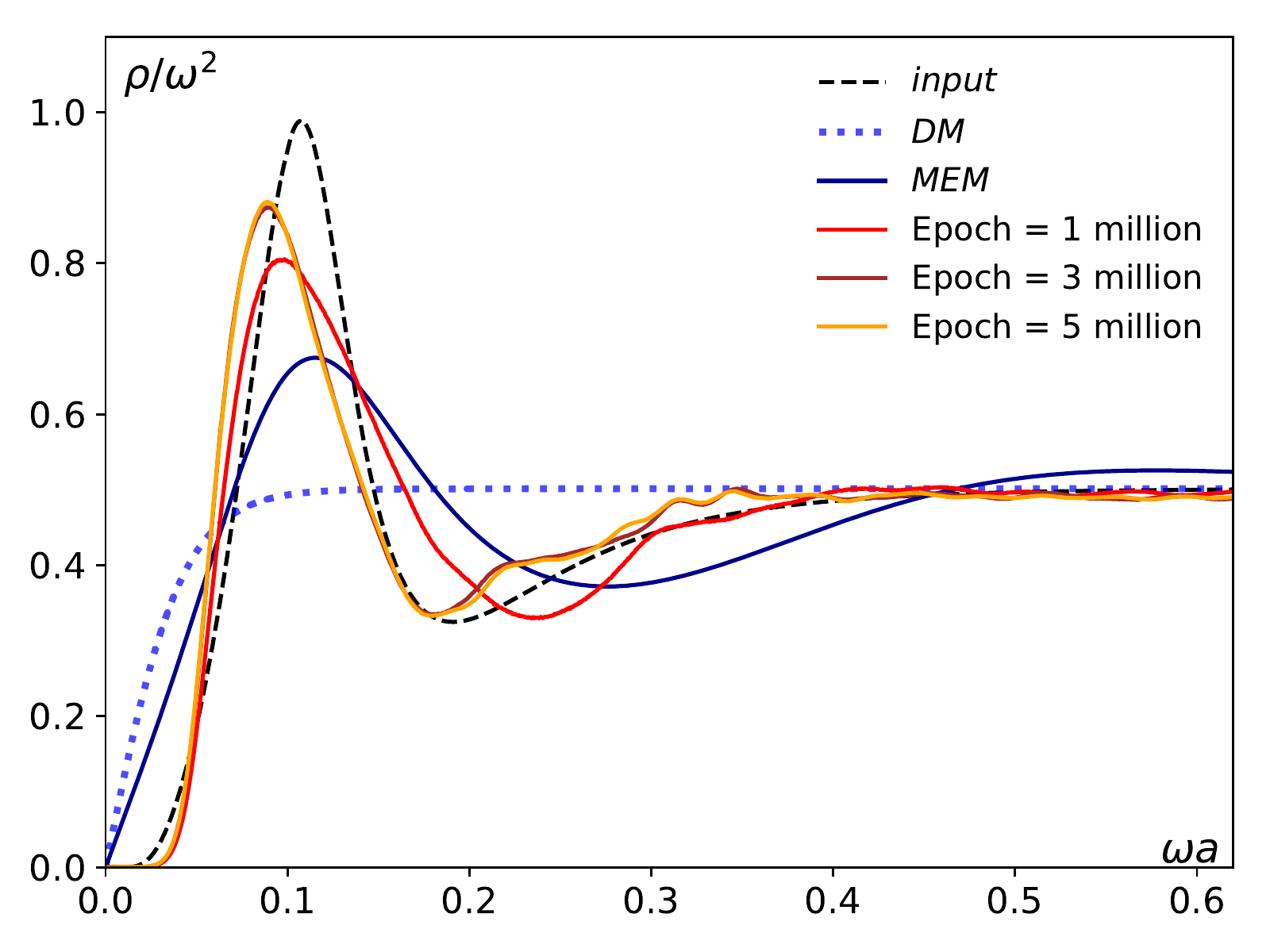}
	\caption{Dependence on hyperparameters and training parameters with $N_\tau=48$. Top left: Dependence on the random seed used in generating the training ensembles. Top right: Dependence on the number of Gaussian peaks used in the training spectral functions. Bottom left: Dependence on the number of layers in the encoders and decoders. Bottom right: Dependence on the number of training epochs.}
	\label{fig:spf-hyperparameters}
\end{figure}

We also show the dependences of output spectral functions on various hyperparameters and training parameters in the SVAE. We find that these dependencies of output spectral functions are marginal. I.e. the features we observed in Figs.~\ref{fig:case3_Nt96} and~\ref{fig:case3_Nt48} are stable and reliable. The details on these dependences are listed as follows: 
\begin{itemize}
\item The dependence on the random seed used in generating training samples: We made five independent runs with five million epochs each using different random seeds and corresponding results are shown in the top left plot of Fig.~\ref{fig:spf-hyperparameters}. 
\item The dependence on the number of resonances used in training, $\hat{N}_g$ [cf. Eq.~\ref{eq:training_rho}]: We have doubled $N_g$ to 100 (i.e. refining the mass step of the resonances by a factor of 2) keeping the training to five million epochs. The corresponding results are shown in the top right plot of Fig.~\ref{fig:spf-hyperparameters}.
\item The dependence on the number of hidden layers: We test two cases with one more and two more hidden layers in both the encoder and decoder, i.e. 1) 3 hidden layers per encoder/decoder, and 2) 4 hidden layers per encoder/decoder. Each added hidden layer contains 250 neurons with a Relu activation function. The number of training epochs is kept as five million. The corresponding results are shown in the bottom left plot of Fig.~\ref{fig:spf-hyperparameters}.
\item The dependence on the number of training epochs: We have trained the SVAE with 1, 3 and 5 million epochs. The corresponding results are shown in the bottom right plot of Fig.~\ref{fig:spf-hyperparameters}.
\end{itemize}

In the above tests all the unmentioned hyperparameters remain fixed.

\section{Dependences on the noise model}
\label{app:noise_model}
\begin{figure}[!hpt]
	\centering
	\includegraphics[width=0.8\textwidth]{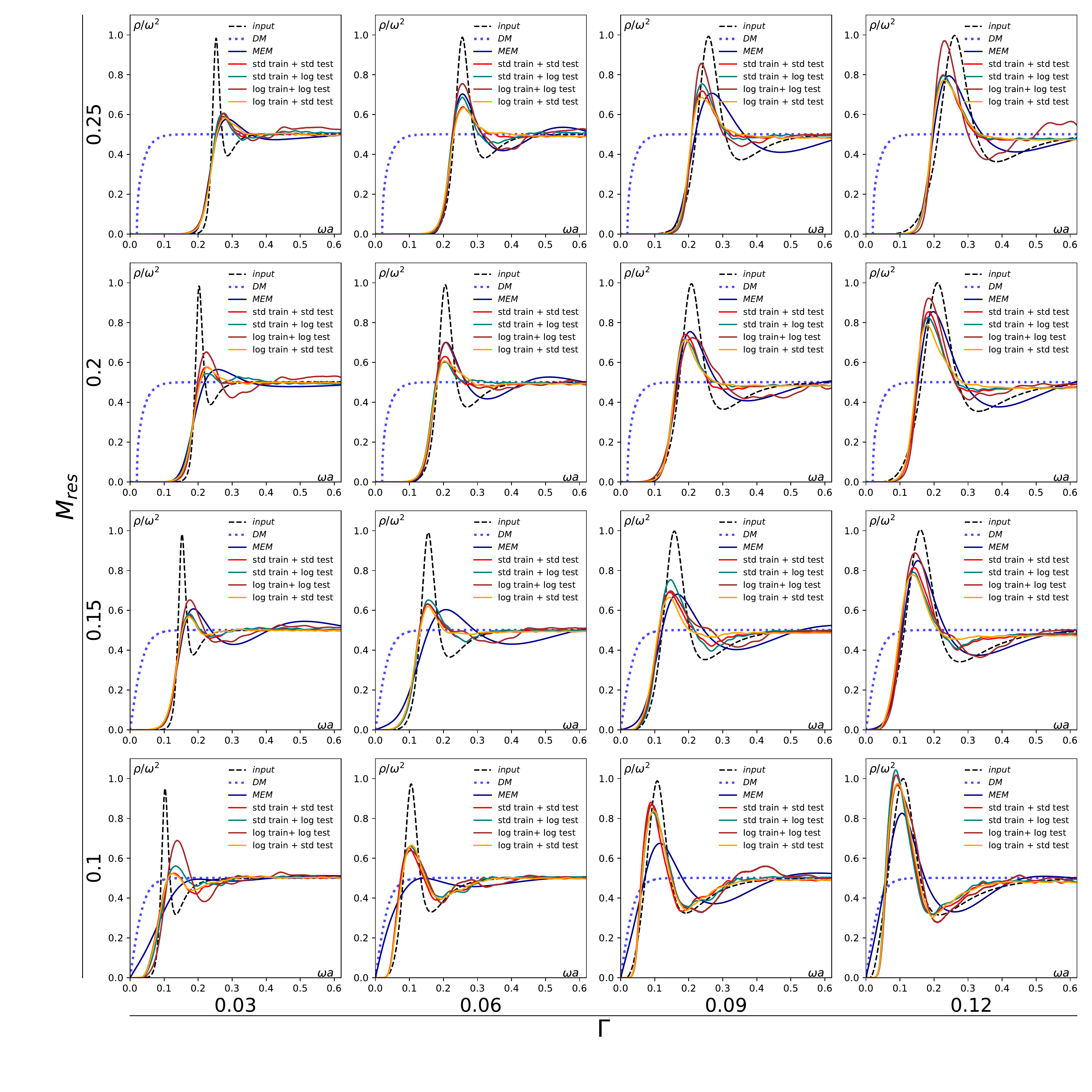}
	\caption{Spectral functions extracted from the mock correlator data with $N_\tau=48$ using two noise models. Here `std' and `log' stand for the noise models where correlators are sampled according to the Gaussian distribution (cf. Eq.~\ref{eq:noise_level}) and the log-normal distribution, respectively. Thus the output spectral functions are obtained using four combinations of noise models in the training (`train') and test (`test'), which are shown as colored solid lines. The input spectral functions as well as the noise levels are the same as those shown in Fig.~\ref{fig:case3_Nt48}. }
	\label{fig:spf-noisemodel}
\end{figure}

\begin{figure}[!hpt]
	\centering
	\includegraphics[width=0.8\textwidth]{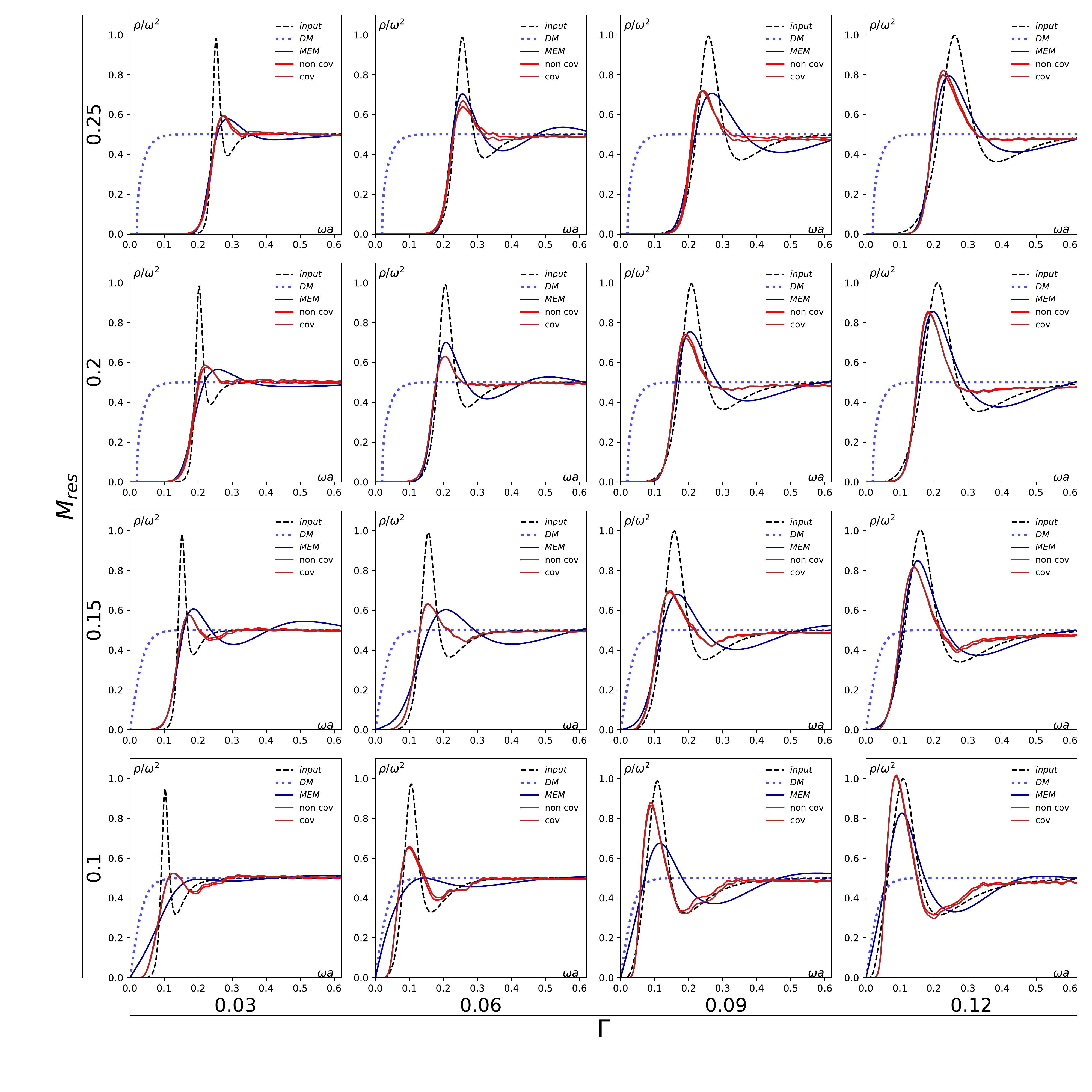}
	\caption{Spectral functions extracted from the test correlator data with $N_\tau=48$ sampled with and without covariance matrix. I.e. the mock correlator data used in the training and test are sampled from the Gaussian distribution and multivariate Gaussian distribution (cf.Eq.~\ref{eq:MG}), respectively. The input spectral functions are the same as those shown in Fig.~\ref{fig:case3_Nt96} and Fig.~\ref{fig:case3_Nt48}.}
	\label{fig:spf-noisemodel-CV}
\end{figure}

In this appendix, we show the test results of spectral functions with various noise models adopted to sample the correlator data used in the training and test processes in Fig.~\ref{fig:spf-noisemodel} and Fig.~\ref{fig:spf-noisemodel-CV}. We find that the mismatch of noise models used in the training and test yields minor differences in the reconstructed spectral functions. 

We tested with following cases
\begin{itemize}
    \item During the training the correlator data are sampled according to a normal distribution (`std', cf. descriptions around Eq.~\ref{eq:noise_level}) and during the test the test correlator data are sampled according to a log-normal distribution (`log'). The results are denoted as ``std train + log test" in Fig.~\ref{fig:spf-noisemodel}.
     \item During the training the correlator data are sampled according to a log-normal distribution, while during the test the correlator are sampled according to a normal or log-normal distribution. The results are denoted as ``log train + std test" and ``log train + log test" in Fig.~\ref{fig:spf-noisemodel} respectively.
     \item Taking into account the correlation among different time slices in test correlators, those test correlators are sampled according to a multivariate Gaussian distribution (cf.~\ref{eq:MG}) with the covariance matrix.
     The covariance matrix is calculated to keep the noise level the same as the lattice QCD data (cf. Eqs.~\ref{eq:cov_d} and~\ref{eq:cov_offd}). The results are shown in Fig.~
     \ref{fig:spf-noisemodel-CV}.
\end{itemize}

Here we also summarize the way used to sample correlator data $G$ during the test and training. We start with a multivariate Gaussian distribution, which can be expressed as
         \begin{equation}
       p(G;G_{gt},\Sigma^{mock}) = \frac{1}{(2\pi)^{N/2}|\Sigma^{mock}|^{1/2}} \exp\left(-\frac{1}{2}\left(G-G_{gt}\right)^T\left(\Sigma^{mock}\right)^{-1}\left(G-G_{gt}\right)\right).
       \label{eq:MG}
    \end{equation}
Here $N$ is number of time slices of correlator used in the computation, and $\Sigma^{mock}$ is the covariance matrix of the mock data and is a $N\times N$ square matrix. When $\Sigma^{mock}$ is a diagonal matrix the above equation goes back to the standard Gaussian distribution.

To mimic the noise level of the lattice correlator, we will need to compute the covariance matrix of the available lattice data $\Sigma^{lat}$, which is defined as
    \begin{equation}
   \Sigma_{ij}^{lat} =\frac{1}{N_c} \sum_{m=1}^{N_c} (G_i^{m,lat} - \bar{G}_i^{lat})(G_j^{m,lat}  - \bar{G}_j^{lat}).
\end{equation}
    Hereafter $i,j$ stand for the time slice of the correlator, and $\bar{G}_{i}^{lat}=\sum_{m=1}^{N_c} G_i^{m,lat}/N_c$ is the average value of the $i$-th time slice of correlator over $N_c$ configurations. We thus rescale 
    $\Sigma^{lat}$ to obtain $\Sigma^{mock}$ to keep the same noise level as in the lattice calculations. 
    For the diagonal part of $\Sigma^{mock}$, we have
    \begin{equation}
        \Sigma_{ii}^{mock} = G_{gt,i}\,\frac{\Sigma_{ii}^{lat}}{\bar{G}_i^{lat}}.
        \label{eq:cov_d}
    \end{equation}
    Here $G_{gt,i}$ is the definite value of the $i$-th time slice of correlator computed from the input mock spectral function $\rho_{mock}$ via Eq.~\ref{eq:Grho}.
    We can then construct the off-diagonal part of $\Sigma^{mock}$, 
    \begin{equation}
        \Sigma_{ij}^{mock} = \Sigma_{ii}^{mock}\, \frac{\Sigma_{ij}^{lat}}{\Sigma_{ii}^{lat}}.
        \label{eq:cov_offd}
    \end{equation}
Once $\Sigma^{mock}$ is obtained, one can then generate $\{G_{i}\}$ according to a multivariate Gaussian distribution, i.e. Eq.~\ref{eq:MG} with $\Sigma^{mock}$.

\section{H\"older's inequality}
\label{app:inequality}
	
In this appendix we show the derivation of Eq.~\eqref{eq:evidence} via the H\"older's inequality~\cite{AnalyticInequalities2013}, which states that
\begin{equation}
  \int\!\! \mathrm{d}\mu\ | f g |  \leq  \left( \int\!\! \mathrm{d}\mu\ |f|^p \right)^{\frac1p} \left( \int\!\! \mathrm{d}\mu\ |g|^q \right)^{\frac1q} 
\end{equation}
if $p,q \geq 1$ and $1/p + 1/q = 1$. Applying the inequality to $P(G|z)$ with Eq.~\eqref{eq:prior} and Eq.~\eqref{eq:likelihood} we arrive at

	\begin{equation}
	\begin{aligned}
	\int  \mathcal{D}\rho(z) \frac{1}{Z_SZ_L} e^{-L+S} &  \leq  \frac{1}{Z_SZ_L}  \left[\int \mathcal{D}\rho(z)e^{ -pL}\right]^{\frac{1}{p}} \left[\int \mathcal{D}\rho(z)e^{qS}\right]^{\frac{1}{q}}\\
	&= \frac{1}{Z_SZ_L} \left[ \int \mathcal{D}\hat{G}[\rho(z)] \left(\frac{\mathcal{D}\hat{G}[\rho(z)]}{\mathcal{D}\rho(z)}\right)^{-1} e^{ -p L}\right]^{\frac{1}{p}} \left[\int \mathcal{D}\rho(z)e^{qS}\right]^{\frac{1}{q}}\\
	&\approx \prod\limits_{j=\tau_{min}}^{N_{\tau}/2} \frac{\Lambda^{\frac{1}{p}}}{Z_SZ_L} \left[\frac{2\pi \alpha^2(z)G^2}{p}\right]^{\frac{1}{2p}}\left(\frac{2\pi}{q}\right)^{\frac{N_{\omega}}{2q}},
	\end{aligned}
	\end{equation}
where $\mathcal{D}x = \prod\limits_{k=1}^{N} dx_k$ and $\Lambda$ is a constant only depending on the kernel $K(\tau, \omega, T)$ (cf. Eq.~\eqref{eq:kernel}):
	\begin{equation}
	\begin{aligned}
	\Lambda = \left(\frac{\mathcal{D}\hat{G}[\rho(z)]}{\mathcal{D}\rho(z)}\right)^{-1} = \left(\frac{\prod\limits_{j=\tau_{min}}^{N_{\tau}/2} d\sum\limits_{l=1}^{N_{\omega}} K_{j,l}(T)\rho^l(z)}{\prod\limits_{l'=1}^{N_{\omega}}d\rho^{l'}(z)}\right)^{-1}=\left( \prod\limits_{l'=1}^{N_{\omega}}\prod\limits_{j=\tau_{min}}^{N_{\tau}/2}\sum\limits_{l=1}^{N_{\omega}} K_{j,l}(T)\delta^l_{l'} \right)^{-1}=\left( \prod\limits_{l'=1}^{N_{\omega}}\prod\limits_{j=\tau_{min}}^{N_{\tau}/2} K_{j,l'}(T) \right)^{-1}.\
	\end{aligned}
	\end{equation}

\end{appendix}
\selectlanguage{english}
\bibliography{ML_spf}
------------
\end{document}